\documentclass[12pt]{article}

\newcommand{\bm}[1]{\mbox{\boldmath$#1$}}

\renewcommand{\baselinestretch}{1.1}
\newcommand{\curl}[0]{{\rm curl}\,}
\usepackage{amsmath}
\usepackage{amssymb}
\usepackage{latexsym}

\begin{document}

\title{General Nonlinear 2-Fluid Hydrodynamics of\\ Complex Fluids and Soft Matter}
\author{H. Pleiner$^{\dagger}$, J.L. Harden$^{\ddagger}$ \\[4mm]
    {\normalsize $^{\dagger}$\it Max-Planck-Institute for  Polymer Research,
55021
      Mainz, Germany}\\
    {\normalsize $^{\ddagger}$\it Chemical Engineering, Johns Hopkins
University,
      Baltimore, MD 21218, USA}
  }
\date{\small appeared in Nonlinear Problems of Continuum Mechanics,  Special issue of  Notices of Universities. South of Russia. Natural sciences,  p.46 - 61 (2003)}
\maketitle

{\abstract{ We discuss general 2-fluid hydrodynamic equations for complex
fluids, where one kind is a simple Newtonian fluid, while the other is either
liquid-crystalline or polymeric/elastomeric, thus being applicable to
lyotropic liquid crystals, polymer solutions, and swollen elastomers. The
procedure can easily be generalized to other
complex fluid solutions. Special emphasis is laid on such
nonlinearities that originate from the 2-fluid description, like the transport
part of the total time derivatives.
It is shown that the proper velocities, with which the hydrodynamic quantities
are convected, cannot be chosen at will, since there are subtle relations
among them. Within allowed combinations the convective velocities are
generally material dependent. The so-called stress
division problem, i.e. how the nematic or elastic stresses are distributed
between the two fluids, is shown to depend partially on the choice of the
convected velocities, but is otherwise also material dependent. A set of
reasonably simplified equations is given as well as a linearized version of an
effective concentration dynamics that may be used for comparison with
experiments.}}

\pagebreak

\section{Introduction \label{intro}}

The thermodynamic and hydrodynamic properties of multi-component complex fluids
are determined by the microscopic degrees of freedom of their constituents and
the coupling between these degrees of freedom.
Such systems can exhibit rather rich phase behavior and dynamics,
especially when one or more components is a structured or macromolecular
fluid~\cite{larsonbook}.
Due in part to the coupling of internal degrees of freedom,
these systems can also exhibit novel flow-induced structural evolution phenomena,
including shear-induced phase transformations and flow alignment of constituents
on microscopic to mesoscopic length scales.
Such structural evolution in turn leads to nonlinear rheological behavior,
such as stress overshoots in response to imposed rates of strain, plasticity, and
 thixotropy.

The overwhelming complexity of the microscopic description of these systems,
such a detailed description is often not well suited for analysis of
the macroscopic dynamical behavior.
Instead,  explicit macroscopic models have been developed for this purpose.
Some such models have been obtained by a suitable coarse-graining procedure
starting
from a microscopic theory.
Others are purely phenomenological models constrained only by conservation laws,
symmetry considerations and thermodynamics.
The so-called ``two-fluid'' models for  binary systems of distinct components
or phases are useful examples of such a macroscopic approach~\cite{reichl}.
In the two-fluid description, each component or phase is treated as a continuum
described by local thermodynamic variables (e.g. temperature, density, and
relevant order parameters), and dynamical quantities (e.g. velocity or momentum).
In general, these variables for the constituents are coupled.  For instance, the
effective friction between components in a binary fluid mixture leads to a drag
force
in the macroscopic description that is proportional to the local velocity
difference.

Two-fluid models have been employed in many different physical contexts.
The two-fluid approach is a key element of many traditional models for
multi-phase
flow of bubbly liquids, fluid suspensions of particulates, and binary mixtures of
simple fluids~\cite{drewbook}.    Other examples in condensed matter physics
include two-fluid models for superfluid helium~\cite{martin},
dynamics of plasmas~\cite{park}, transport in superconductors~\cite{lee},
viscoelasticity of concentrated fluid emulsions~\cite{hebraud},
flow-induced ordering of wormlike micelle solutions~\cite{kadoma},
flow of colloidal suspensions~\cite{lhuillier}.
Two-fluid models have been used extensively to model a wide range of
dynamical phenomena in polymer solutions and binary blends,
including the hydrodynamics modes of quiescent polymer solutions~\cite{brochard1,
harden},
kinetics of polymer dissolution~\cite{brochard2},  hydrodynamics and rheology
of polymer solutions and blends~\cite{doi1}-\cite{hashimoto},
and polymer migration and phase separation under flow~\cite{doi2}-\cite{beris1}.

These examples share certain general features.  In each, two distinct species or
coexisting phases
(gas and liquid, normal fluid and superfluid, polymer and solvent, mesogens and
solvent etc.)
with mass densities $\rho_{1}$ and $\rho_{2}$,
which are conserved individually in the absence of chemical reactions,
move with distinct velocities $\bm{v}_{1}$ and $\bm{v}_{2}$, respectively.
Due to (usually strong) internal friction, the momenta of the constituent species,
$\rho_{1} \bm{v}_{1}$ and $\rho_{2} \bm{v}_{2}$, are not conserved individually.
Of course, total momentum is conserved. In most cases of fluid mixtures the
friction
is so strong that the velocity difference $\bm{v}_{1} - \bm{v}_{2}$ is nonzero
for very
short times only, i.e.~it is a very rapidly relaxing quantity that is not included
in the hydrodynamic description for binary mixtures.
However, there are systems and situations, where the relaxation of
the relative momenta is slow enough to have a significant influence even on the
hydrodynamic time scale. Then a two-fluid description is appropriate and useful.

In this communication we focus on a general nonlinear two-fluid description of
complex fluids, where one species is a viscous Newtonian fluid and the other
either a polymer or a liquid crystal. Emphasis is placed on the rigorous
derivation of the dynamic equations within the framework of hydrodynamics as
contrasted to ad-hoc treatments. The resulting equations are rather general
and complicated. They can and have to be simplified for special applications
or systems by appropriate and well-defined approximations. One of the
advantages of starting from the general theory is the possibility to identify
and characterize the approximations made. The hydrodynamic method,
described in some detail in \cite{mbuch}-\cite{Forster}, is quite general and
rigorous,
being based on symmetries, conservation laws, and thermodynamics.
In the following sections, we provide a detailed analysis of two-fluid models
for lyotropic nematogens in a simple viscous solvent, followed by an abbreviated
extension of this treatment for isotropic elastomers (e.g. entangled polymer
solutions and gels)
in a simple viscous solvent.  We close with a discussion of our general results
and
their possible implications for experiments.

\section{Thermodynamics \label{princ}}

The hydrodynamics of fluid mixtures as described above is governed by
conservation laws (individual masses, total momentum and total energy),
balance equations for the liquid crystalline degrees of freedom, for the
transient
elasticity of polymers and for the relaxation of relative momentum. There are
different ways of writing the appropriate equations. One popular choice is to
use equations for
individual mass densities and individual momentum densities, another to use
the mass density and one concentration variable and the total momentum density
and the relative velocity difference. Since they both have their advantages
and disadvantages we will present both ways of description and show, how they
are connected. In this and the following sections we will use a nematic liquid
crystal as the second, complex fluid. Transcription of the formulas to the
polymer
case will be given in Sec.(\ref{polymers}).

The starting point of any macroscopic description is the total energy $E$ of
the
system as a function of all the relevant variables. Since the energy is a
first
order Eulerian form of the extensive quantities, we can write
\begin{equation}
 \label{energy}
E= \epsilon \,V = \int \epsilon\, dV = E(M_1,\, M_2,\, V,\, \bm{G_1},\,
\bm{G_2},
\, S,\, M_2
\nabla_j n_i,\, M_2 \delta n_i)
\end{equation}
The masses, $M_1,\,M_2$ and momenta $\bm{G_1},\,\bm{G_2}$ of species
1 and 2 are related to the appropriate (volume) densities by $\rho_1 = M_1 /V,
\,
\rho_2 =
M_2 / V,\, \bm{g}_1 = \bm{G}_1 /V = \rho_1 \bm{v}_1,\, \bm{g}_2 = \bm{G}_2 /V
=
\rho_2 \bm{v}_2$, while for the entropy density $\sigma = S/V$. The nematic
degrees of freedom are related to species 2 and consist of director rotations
$\delta n_i$. The nematic phase shows orientational order along the line
denoted
by $\bm{n}$ (with $\bm{n}^2 =1$) called the director. Since up and down (along
that line) cannot be discriminated, all equations have to be invariant
under a $\bm{n} \to -\bm{n}$ transformation. Homogeneous rotations do not cost
energy, so in a linear description (of the field-free case) $\delta n_i$ is
absent in $E$ and only gradients $\nabla_j n_i$ enter \cite{deGennes}.
We have kept both terms
to cope with the general case.

Introducing thermodynamic derivatives (partial derivatives where all other
variables are kept fixed) we define temperature $T$, thermodynamic pressure
$p$,
chemical potentials
$\mu_1,\,\mu_2$ and velocities $\bm{v}_1,\,\bm{v}_2$ of the two fluids, as
well
as the conjugate fields $\chi_{ij}$ and $k_i$ connected to the nematic degrees
of freedom
\begin{eqnarray}
T = \frac{\partial E}{\partial S} = \frac{\partial \epsilon}{\partial \sigma},
\qquad \quad \mu_1 =
\frac{\partial E}{\partial M_1} = \frac{\partial \epsilon}{\partial \rho_1},
\quad
\quad \mu_2 =
\frac{\partial E}{\partial M_2} = \frac{\partial \epsilon}{\partial \rho_2}
\nonumber \\
\label{partial}  p = - \frac{\partial E}{\partial V}, \quad \quad
\bm{v}_1 = \frac{\partial E}{\partial \bm{G}_1} = \frac{\partial \epsilon}
{\partial \bm{g}_1}, \quad \quad \bm{v}_2 = \frac{\partial E}{\partial
\bm{G}_2}
= \frac{\partial \epsilon}
{\partial \bm{g}_2} \\  k_i = \frac{\partial E}{\partial (M_2 n_i)} =
\frac{\partial \epsilon}
{\partial (\rho_2 n_i)}, \quad \quad  \chi_{ij} = \frac{\partial E}{\partial
(M_2 \nabla_j n_i)} =
\frac{\partial \epsilon}
{\partial (\rho_2 \nabla_j n_i)} \nonumber
\end{eqnarray}
Expanding eq.(\ref{energy}) into first order differentials, the condition
$dV=0$ leads to an expression for the pressure
\begin{equation}
 \label{pressure}
p= -\epsilon + T \sigma + \rho_1 \mu_1 + \rho_2 \bar \mu_2 + \bm{v}_1 \cdot
\bm{g}_1 + \bm{v}_2 \cdot \bm{g}_2
\end{equation}
where we have introduced the effective chemical potential of the nematic $\bar
\mu_2 = \mu_2 + \chi_{ij} \nabla_j n_i + k_i \delta n_i$. In addition, the
differentials are related by the Gibbs relation
\begin{equation}
 \label{Gibbs}
d\epsilon = T d\sigma + \mu_1 \,d\rho_1 + \bar \mu_2 \,d\rho_2 + \bm{v}_1
\cdot
d\bm{g}_1 + \bm{v}_2 \cdot d\bm{g}_2  + \Psi_{ij} \,d \nabla_j n_i + h_i \,dn_i
\end{equation}
with the more familiar nematic conjugate fields $\Psi_{ij} = \rho_2 \chi_{ij}$
and $h_i = \rho_2 k_i$. From eqs.(\ref{pressure}, \ref{Gibbs}) the expression
for the differential pressure results (Gibbs-Duhem relation) that is useful in
switching from pressure to chemical potentials or vice versa
\begin{equation}
 \label{Duhem}
dp = \sigma \,dT + \rho_1 \,d\mu_1 + \rho_2 \,d\bar \mu_2
+ \bm{g}_1 \cdot d\bm{v}_1 + \bm{g}_2 \cdot d\bm{v}_2 - \Psi_{ij} \,d \nabla_j
n_i - h_i \,dn_i
\end{equation}

A second set of equations is obtained by switching to the total density,
$\rho = \rho_1 + \rho_2$, and the total momentum, $\bm{g} = \bm{g}_1 +
\bm{g}_2 = \rho_1 \bm{v}_1 + \rho_2 \bm{v}_2$, which are the sums of the
original quantities and which are both conserved
quantities. The two-fluid nature has then to be
represented by additional variables. A natural choice seems to be the use of
the
density and momentum differences. However the latter choice is problematic,
since it necessarily implies the conjugate quantities also to be the
(arithmetic) sums and
differences of the original conjugate quantities. Thus, the conjugate to
$\bm{g}$ would be $\bm{v_{1}} + \bm{v_{2}}$, which does not reflect correctly
the possible one-fluid limits $\rho_{1} \to 0$ or $\rho_{2} \to 0$. The
physically acceptable conjugate to the total momentum is the mean velocity
$\bm{v}$
defined by $\rho^{-1} \bm{g}$. Insisting on $\bm{v}$,
the mean velocity, to be the conjugate of the
total momentum $\bm{g}$, the choice of the remaining variable describing the
different velocities is severely limited. Compatibility with (\ref{Gibbs})
allows as variable only the velocity difference
*** $\bm{w} \equiv \bm{v_{1}} -
\bm{v_{2}}$ (with $\bm{m} \equiv \rho^{-1} \rho_{1}
\rho_{2} \bm{w}$ as conjugate quantity) or more generally $\alpha
 \bm{w}$ as variable with
$\alpha^{-1}\rho_{1}\rho_{2}\rho^{-1} \bm{w}$ as conjugate,
where $\alpha$ can be freely choosen. There is no a-priori advantage for any
of the choices and we will stick to $\alpha= 1$.\footnote{The choice $\alpha =
  \rho_{1}\rho_{2}\rho^{-1}$ would just interchange the roles of $\bm{w}$ and
  $\bm{m}$ as variable and conjugate.}
*** From $\bm{w}= \bm{g}_1 /
\rho_1 - \bm{g}_2 / \rho_2$
one gets
\begin{equation}
\label{v12}
\bm{v}_1 = {\rho}^{-1} \bm{g} + (1 - \phi) \bm{w}, \quad \quad  \bm{v}_2 =
{\rho}^{-1} \bm{g} - \phi \,\bm{w}
\end{equation}
The representation of the two different densities is less problematic. A
convenient choice for that variable is the concentration, $\phi =
\rho_1 / \rho$, with
$\rho_2/\rho = 1 - \phi$.
If the expansion coefficients of the two fluids are the same,
$\phi$ can be interpreted as the volume fraction as well.
Instead of $\phi$ one could have used, e.g. the density difference $\rho_{1} -
\rho_{2}$ (or any other linear combination of $\rho_{1}$ and $\rho_{2}$
different from $\rho$) as variable without much change.

After some trivial
algebra eqs.(\ref{pressure}-\ref{Duhem}) can be
written in the new variables as
\begin{eqnarray}
 \label{pressure2}
p&=& -\epsilon + T \sigma + \rho \mu + {\rho}^{-1} \bm{g}^{\,2} \\
\label{Gibbs2}
d\epsilon &=& T\, d\sigma + \Pi^\prime\, d\phi +  \mu\, d\rho + \bm{v} \cdot
d\bm{g}  + \bm{m} \cdot d\bm{w} + \Psi_{ij}\, d \nabla_j n_i + h_i\, dn_i \\
\label{Duhem2}
dp &=& \sigma\, dT + \rho\, d\mu
+ \bm{g} \cdot d\bm{v} - \bm{m} \cdot d\bm{w} - \Pi^\prime\, d\phi  -
\Psi_{ij}\,
d \nabla_j n_i - h_i\, dn_i
\end{eqnarray}
where we have introduced the relative pressure $\Pi^\prime$, the total
chemical
potential $\mu$, the mean velocity $\bm{v}$ and the weighted relative momentum
$\bm{m}$ defined by
\begin{eqnarray}
\nonumber
\Pi^\prime &=& \rho\,(\mu_1  - \bar \mu_2)  + \bm{w} \cdot \bm{g} + \rho\,
\bm{w}^2
(1-2\phi) \equiv \rho \, \Pi \\ \nonumber
\mu &=& \mu_1 \phi + \bar \mu_2 (1 - \phi)  + \bm{w}^2 \phi (1 - \phi) \\
\label{conjugates2}
{\rm or\,\,\, vice \,\,\, versa} \quad \quad
\mu_{1} &=& \mu + \rho^{-1} \rho_{2}\, (\Pi - \bm{w}\cdot \bm{v}_{1}) \\
\nonumber
\mu_{2} &=& \mu - \rho^{-1} \rho_{1}\, (\Pi + \bm{w} \cdot \bm{v}_{2}) \\
\nonumber
{\rm where} \quad\quad
\bm{v} &=&  \phi \,\bm{v}_1 + (1 - \phi) \,\bm{v}_2 = \rho^{-1}(\bm{g_{1}} +
\bm{g_{2}})\\
\nonumber
\bm{m} &=& \rho \,(1- \phi) \phi \,\bm{w} = ({\rho_2}\bm{g}_1 -
{\rho_1} \bm{g}_2) \rho^{-1}
\end{eqnarray}
The Gibbs relations connects variables that show different rotational behavior.
Energy, entropy, the densities and the concentration are scalar quantities that
do not change under (rigid) rotations, i.e. $d\epsilon = d\sigma = d\rho =
d\rho_1
=d\rho_2 = d\phi =0$. The vectors and tensors are transformed according to $d n_i
 =
\Omega_{ij} n_j, \, d g_i =  \Omega_{ij} g_j, \, d w_i =\Omega_{ij} w_j , \,
d \nabla_j n_i =\Omega_{jk} \nabla_k n_i  + \nabla_j  \Omega_{ik} n_k$, where
$\Omega_{ij}=- \Omega_{ji}$ is any constant antisymmetric matrix. The rotational
invariance of the Gibbs relation (\ref{Gibbs},\ref{Gibbs2}) then leads to the
relation
\begin{equation}
\label{rotcond}
h_i n_j + \Psi_{ki} \nabla_j n_k + \Psi_{ik} \nabla_k n_j =
h_j n_i + \Psi_{kj} \nabla_i n_k + \Psi_{jk} \nabla_k n_i
\end{equation}
which has to be fulfilled by the conjugate quantities. There are no contributions
from the momenta and velocities, since $\bm{g} \parallel \bm{v}$, $\bm{w}
\parallel
\bm{m}$, and $\bm{g_{1,2}} \parallel \bm{v_{1,2}}$. Relation (\ref{rotcond}) is
useful for reformulating the stress tensor, in particular to symmetrize it
explicitly
\cite{mpp}.

Having set up the thermodynamics of the relevant variables we are now in a
position to establish the structure of the dynamic equations.
\\[0.5cm]

\section{Dynamic Equations \label{equations}}

For the two fluids there are independent continuity equations stating that
neither mass can be destroyed nor created, but only transported. Transport can
involve convection as well as (relative) diffusion.
This leads immediately to
\begin{eqnarray}
\label{contin1}
\dot \rho_1 + \nabla_i (\rho_1 v_i^{\,(1)} + j_i^{\,(1)}) = 0 \\
\label{contin2}
\dot \rho_2 + \nabla_i (\rho_2 v_i^{\,(2)} - j_i^{\,(1)}) = 0
\end{eqnarray}
When dealing with components of vectors, the subscripts $1,\,2$ are
written as superscripts for clarity. The phenomenological mass currents in
Eqs.(\ref{contin1}, \ref{contin2}) add up to zero, since the total mass current
is equal to the total momentum density $\bm{g} = \rho \,\bm{v}$.
Eqs.(\ref{contin1}, \ref{contin2}) can be rewritten in terms of the total
density
 and the concentration as
\begin{eqnarray}
\label{contin}
\dot \rho + \nabla_j \rho \, v_j &=& 0 \\ \label{concdot}
\dot \phi + v_j \nabla_j \phi +  \rho^{-1} \nabla_i  \left( \rho \phi (1-\phi)
  w_i+ j_i^{\,(1)}
\right)  &=& 0
\end{eqnarray}
 show the
characteristic
difference between extensive quantities, where convection is of the form
$\bm{\nabla} \cdot (\bm{v} *)$
and intensive ones with $\bm{v} \cdot \bm{\nabla} *$.

Note that the concentration does not obey
a conservation law, except when linearized around a zero-velocity state or if
$\rho = const.$ is assumed. Because the mass current density of the total
fluid is equal to the momentum density $\bm{g}$ ($= \rho\bm{v}$), the total
mass is convected by the mean velocity in (\ref{contin}). In
Eqs. (\ref{contin1}, \ref{contin2}, \ref{concdot}) the convective terms are
not fixed a
priori, since the phenomenological current $\bm{j}_{1}^{\,(1)}$ can contain
contributions proportional to some velocities, thus altering the effective
velocity, with which the different quantities are convected. We will discuss
this point extensively after having derived the full set of equations.

The dynamic equations for the other variables are even more complicated and
also contain phenomenological parts. These are expressed by yet to be
determined
currents. But they also contain convective (or transport) terms. Therefore, we
can set up the following equations as an ansatz
\begin{eqnarray}
\label{epsdot}
\dot \epsilon + \nabla_j (\epsilon + p) v_j  + \nabla_i j_i^{\,(\epsilon)} &=&
 0
\\
\label{sigmadot}
\dot \sigma + \nabla_j \sigma v_j + \nabla_i j_i^{\,(\sigma)} &=&  R/T \\
\label{gdot}
\dot g_i + \nabla_j g_i v_j + \nabla_j \sigma_{ij} &=& 0  \\
\label{wdot}
\dot w_i + v_j \nabla_j w_i + X_i  &=& 0 \\ \label{ndot}
\dot n_i + v_j \nabla_j n_i + Y_i  &=& 0
\end{eqnarray}
containing either the divergence of a current
($j_i^{\,(\epsilon)}, \, j_i^{\,(\sigma)},\, j_i^{\,(1)}, \,
\sigma_{ij}$) when conservation laws are involved, or
quasi-currents ($X_i,\, Y_i$) in the case of balance equations for
non-conserved variables. Each of the currents and quasi-currents
consists generally of three parts: A geometric or symmetry related
one without any phenomenological coefficients, which we will
determine below, and two phenomenological parts, which are either
reversible (superscript $rev$) or irreversible (superscript
$dis$). The phenomenological parts will be discussed in section
\ref{dynamics}. The entropy balance (\ref{sigmadot}) is not a
conservation law, since for irreversible processes the entropy
production $R$ has to be positive and only for purely reversible
actions $R=0$. In Eqs. (\ref{contin}-- \ref{ndot}) the convective
terms are written down such that all quantities are convected by
the {\em same} velocity.\footnote{In the energy conservation law
(\ref{epsdot}) the free enthalpy $\epsilon + p$ is convected,
cf.\cite{Forster}} This is dictated by the postulation of zero
entropy production (these transport terms are reversible).
However, it should be repeated that the phenomenological
reversible currents may change the effective convection velocity,
something we will discuss later.

Putting the dynamic equations
(\ref{contin}--\ref{ndot}) into the Gibbs relation (\ref{Gibbs}) the condition
$R=0$ ($R>0$) for the convective and the reversible (dissipative)
phenomenological parts of the currents, leads to the following conditions
\begin{eqnarray}
\label{sigma}
\sigma_{ij} &=& \delta_{ij} p + \Psi_{kj} \nabla_i n_k  +
\sigma_{ij}^{\,(rev)} +
\sigma_{ij}^{\,(dis)} \\
\label{X}
X_i &=& \nabla_i \Pi + X_i^{\,(rev)} + X_i^{\,(dis)} \\
\label{Y}
Y_i &=&
Y_i^{\,(rev)} + Y_k^{\,(dis)}\\
\label{jsigma}
j_i^{\,(\sigma)} &=& j_i^{\,(\sigma,rev)} + j_i^{\,(\sigma,dis)}\\
\label{jphi}
j_i^{\,(1)} &=& j_i^{\,(1,rev)} + j_i^{\,(1,dis)}
\end{eqnarray}
with the generalized conjugate to the nematic degrees of freedom $\bar h_i =
h_i -
 \nabla_j \Psi_{ij} = \rho_2 (k_i - \nabla_j \chi_{ij})$.
The stress tensor $\sigma_{ij}$ contains the isotropic pressure $p$
(\ref{pressure2}), while the
quasi-current $X_i$ of the relative velocity contains the gradient of $\Pi$,
the
relative pressure divided by the total density, (\ref{conjugates2}).
The terms related to the nematic
degrees of freedom are well-known from ordinary nematodynamics.
The energy conservation law is redundant here, because
of the Gibbs relation (\ref{Gibbs}) and $j_{i}^{\,(\epsilon)}$ is not needed.

The phenomenological parts have to fulfill (up to an irrelevant divergence
term)\footnote{The true condition is $\int R \,dV \geq 0$.}
\begin{equation}
\label{Rcond}
R = -j_i^{\,(\sigma,*)} \nabla_i T - j_i^{\,(1,*)} \nabla_i \Pi - \sigma_{ij}^
{\,(*)} \nabla_j
v_i + \bar h_i \,
Y_i^{\,(*)} + m_i \, X_i^{\,(*)} \geq 0
\end{equation}
with the equal sign ($>$ sign) for $* = rev$ ($*=dis$), respectively.

Eq.(\ref{Rcond}) also reveals the equilibrium conditions
\begin{equation} \label{Equicond}
\nabla_{i} T=0 \quad\quad\quad \nabla_i \Pi =0 \quad\quad\quad A_{ij}=0 \quad
\quad\quad
\bar h_{i}=0
\quad\quad\quad m_{i}=0
\end{equation}
where $2A_{ij}= \nabla_{j}v_{i} + \nabla_{i} v_{j}$.

Before we will determine the
phenomenological parts in (\ref{sigma}--\ref{jsigma}),
we first have a look into the 2-fluid statics.
\\[0.5cm]

\section{Statics \label{statics}}

The statics is given by the connection of thermodynamic conjugates with the
variables. The conjugates are defined by partial derivatives of the energy
density (\ref{partial}). Thus one can either write down a phenomenological
energy expression and take the derivatives or give directly these
relations under the proviso that mixed derivatives are equal. Two of these
connections have already been given in eq.(\ref{conjugates2}) relating
$\bm{v}$ with ${\bm{g}}$ and $\bm{m}$ with $\bm{w}$. Of course, these are not
really static relations. They are fixed (and not of phenomenological nature),
since the mass current $\rho \,\bm{v}$ is identical to the momentum density and
since the kinetic energy density is $(1/2)\rho_{1} \bm{v}_{1}^{2} +
(1/2)\rho_{2} \bm{v}_{2}^{2}$.

The 3 scalar conjugates \{$T,\,\Pi,\, \mu$\}
have to be expressed by the variables \{$\sigma,\,\phi,\,\rho$\} or using the
other set of variables \{$\sigma,\,\rho_{1},\,\rho_{2}$\} and conjugates
\{$T,\, \mu_{1},\,\mu_{2}$\} by
\begin{eqnarray}
\label{susT}
\delta T &=& \frac{T}{C_{V}}\delta \sigma + \frac{1}{\rho \alpha_{1}}
\delta \rho_{1} + \frac{1}{\rho \alpha_{2}}
   \delta \rho_{2}  \\
\label{susmu1}
\mu_{1} &=& \frac{1}{\rho^{2} \kappa_{1}} \delta \rho_{1}+
\frac{1}{\rho^{2} \kappa_{3}} \delta \rho_{2} +
\frac{1}{\rho \alpha_{1}} \delta \sigma \\
\label{susmu2}
\bar \mu_{2} &=& \frac{1}{\rho^{2} \kappa_{2}} \delta \rho_{2}+
\frac{1}{\rho^{2} \kappa_{3}} \delta \rho_{1} +
\frac{1}{\rho \alpha_{2}} \delta \sigma
\end{eqnarray}
The other conjugates $\Pi$ and $\mu$ have been related to $\mu_{1}$ and $\bar
\mu_{2}$ in (\ref{conjugates2}) and are therefore also fully
determined
\begin{eqnarray}
\label{susT2}
\delta T &=&   \frac{T}{C_{V}}\delta \sigma +  \frac{1}{\alpha_{\phi}} \delta
\phi + \frac{1}{\rho \alpha_{\rho}}  \delta \rho \\
\label{suspi}
\Pi &=& \frac{1}{\rho \kappa_{\phi}} \delta \phi +
\frac{1}{\rho^2 \kappa_{\pi}} \delta \rho +
\frac{1}{\rho \alpha_{\phi}} \delta \sigma  + \bm{w}\cdot \bm{v} +
  \bm{w}^{2}(1-2\phi) \\
\label{susmu}
\mu &=& \frac{1}{\rho^{2} \kappa_{\mu}} \delta \rho +
\frac{1}{\rho \kappa_{\pi}} \delta \phi +
\frac{1}{\rho \alpha_{\rho}} \delta \sigma + \bm{w}^{2} \phi (1-\phi)
\end{eqnarray}
with
\begin{eqnarray} \label{alpha1}
{\alpha_{\phi}}^{-1} &=& {\alpha_{1}}^{-1} -
  {\alpha_{2}}^{-1} \\
{\alpha_{\rho}}^{-1} &=& {\phi}{\alpha_{1}}^{-1} +
(1-\phi){\alpha_{2}}^{-1} \\
{\kappa_{\phi}}^{-1} &=& {\kappa_{1}}^{-1} +
{\kappa_{2}}^{-1} - 2 {\kappa_{3}}^{-1} \\
{\kappa_{\pi}} ^{-1} &=& {\phi}{\kappa_{\phi}}^{-1} -
{\kappa_{2}}^{-1} + {\kappa_{3}}^{-1} \\
{\kappa_{\mu}}^{-1} &=& {\phi^{2}}{\kappa_{1}}^{-1} -
{(1-\phi)^{2}}{\kappa_{2}}^{-1} \label{alpha2}
\end{eqnarray}
Eqs.(\ref{susT}--\ref{susmu2}) as well as (\ref{susT2}--\ref{susmu})
contain 6 static susceptibilities as
compared to 3 in a 1-fluid description. In addition to the specific heat
$C_{V}$ there are 2 thermal expansion coefficients (since there are 2
densities) and 3 compressibilities (2 diagonal and one cross term).
Eqs.(\ref{susT}--\ref{susmu2})
are linear in the deviations from equilibrium,
while (\ref{susT2}--\ref{susmu}) explicitly contains nonlinear corrections
involving velocities. Of course,
the coefficients can still be phenomenological functions of the scalar
variables (e.g. $T$ or $\sigma$,
$p$ or $\rho$, $\rho_{1}$ and $\rho_{2}$ and even $\bm{w}^{2}$)
giving rise to additional nonlinearities that come with (usually) small
coefficients. Note that neglecting cross-susceptibilities either in
(\ref{susT}--\ref{susmu2}) or in (\ref{susT2}--\ref{susmu}) denotes two
physically distinct (and incompatible) approximations, the justification of
either one is not obvious a priori.

Of course, there are situations where one has to go beyond the approximation
used in the static equations above. Describing spinodal decomposition of the
fluids,
e.g. by
an energy density $\epsilon = a \,\phi^2 + b\,[(1-\phi) \ln (1-\phi) + \chi
\phi
(1-\phi)] + c \,\phi^4 + d\, (\nabla_1 \phi)^2$ immediately
leads to nonlinear and gradient terms w.r.t. $\phi$.

What is left is the determination of $\bar h_{i}=h_{i}-\nabla_{j}\Psi_{ij}$ in
terms of $n_{i}$; cross-couplings to other variables are not possible due to
symmetry. Thus this part of the statics is identical to that of
ordinary nematics and can be taken over without any change
\begin{eqnarray}
 \label{Frank}
\Psi_{ij} &=& K_{ijkl} \,\nabla_{l} n_{k} \\ \label{Franknl}
h_{i}    &=& \delta_{iq}^{\perp}\, \frac{\partial K_{pjkl}}{2\partial
    n_{q}}\,(\nabla_{j}n_{p})(\nabla_{l}n_{k}) -\chi_{a} (\bm{H \!\cdot \!n})
H_{i}
  -\epsilon_{a} (\bm{E \!\cdot \!n}) E_{i}
\end{eqnarray}
with $K_{ijkl} = K_{1}\,\delta_{ij}^{\perp}\, \delta_{kl}^{\perp} +
K_{2}\,n_{p}\, \epsilon_{pij}\, n_{q}\, \epsilon_{qkl} + K_{3}\, n_{j}\, n_{l}
 \,
\delta_{ik}^{\perp}$, the Frank gradient energy, and the transverse Kronecker
symbol, $\delta_{ij}^\perp = \delta_{ij} - n_i n_j$. Orientation effects due
to static external magnetic and electric effects enter through the diamagnetic
($\chi_{a}$) and dielectric ($\epsilon_{a}$) anisotropy. For positive
anisotropies the director is parallel to the external magnetic
or electric field in equilibrium, which leads to a restoring torque outside
equilibrium, e.g. to a (linearized) contribution to $h_{i} =
\chi_{a} \bm{H}^{2} \delta n_{i}$ (with $n_{i}\delta n_{i} =0$). For negative
anisotropies the director is perpendicular to the external fields and
e.g. $h_{i} = |\chi_{a}| (\bm{H} \cdot \delta \bm{n})H_{i} $.

Since $\Psi_{ij}$ and $h_i$  are
proportional to $\rho_2$, so are the $K_n$'s (and $\chi_{a},\epsilon_{a}$).
Again (\ref{Frank}) is linear
in the deviations from equilibrium, but the inherent dependence of the
material tensor on the
direction $\bm{n}$ leads to nonlinearities in (\ref{Franknl}).
\\[0.5cm]

\section{Phenomenological Part of the Dynamics \label{dynamics}}

We now close our system of equations by setting up the connection
between the currents and the thermodynamic conjugates (or rather
their gradients usually called thermodynamic forces). For the
irreversible part this is done by writing the entropy production
in terms of the forces
\begin{eqnarray}
 \label{dissipation}
2R &=& \kappa_{ij} (\nabla_i T)(\nabla_j T) + D_{ij}
(\nabla_i \Pi)
(\nabla_j \Pi)
+ 2 D_{ij}^{\,(T)}(\nabla_i \Pi)(\nabla_j T) + \gamma_1^{-1} \delta_{ij}^\perp
\,
\bar h_i \,\bar h_j \nonumber  \\  &+&\nu_{ijkl} (\nabla_j
 v_i)
(\nabla_l v_k)
 +  \xi_{ij}^\prime m_i m_j
\end{eqnarray}
including heat conduction, diffusion and thermodiffusion
($\kappa_{ij},\, D_{ij},\, D_{ij}^{\,(T)}$, respectively, all of
the form $\kappa_{ij} = \kappa_\perp \delta_{ij}^\perp +
\kappa_\parallel n_i n_j$), director orientational viscosity
$\gamma_1$, and viscosity related to gradients of the mean
velocity $\nu_{ijkl}$. The latter has a $\nu_{ijkl} = \nu_{klij}$
symmetry and is of the uniaxial form \cite{mbuch} characteristic
for nematic systems. The last term in (\ref{dissipation})
describes the mutual friction between the two species as will
become clear below. In (\ref{dissipation} we have neglected
viscosity-like contributions involving the relative velocity
$\nabla_j m_i$, since there is already dissipation due to $m_i$. A
more complete discussion of viscosity in a 2-fluid discussion is
given in the Appendix. The dissipation function given above is
bilinear in the forces, an approximation commonly called linear
irreversible thermodynamics. Nevertheless it leads to
nonlinearities due to (implicit and explicit) dependences of
transport tensors on the variables.

According to (\ref{Rcond}) the dissipative parts of the phenomenological
currents
then follow from differentiating $R$
\begin{eqnarray}
\label{sigmadiss}
j_i^{\,(\sigma,dis)} &=& -(\partial R) / (\partial \nabla_i T) = -\kappa_{ij}
\nabla_j T - \rho \,\phi (1-\phi) \,d_{ij}^{\,(T)} \nabla_j \Pi \\
\label{ndiss} Y_i^{\,(dis)} &=& (\partial R) / (\partial \bar h_i)
= \gamma_1^{-1} \delta_{ij}^\perp \,\bar h_j \\ \label{gdiss}
\sigma_{ij}^{\,(dis)} &=& - (\partial R) / (\partial \nabla_j v_i)
= - \nu_{ijkl}\, \nabla_l v_k \\ \label{wdiss} X_i^{\,(dis)}
&=& (\partial R) / (\partial m_i) = \xi_{ij}^\prime \,m_j  \\
\label{phidiss} j_i^{\,(1,dis)} &=& - (\partial R) / (\partial
\nabla_i \Pi) = -\rho \,d_{ij} \nabla_j \Pi -   \rho \,\phi
(1-\phi) \,d_{ij}^{\,(T)} \nabla_j T
\end{eqnarray}
where we have introduced the usual form of the diffusion ($D_{ij}= \rho
\,d_{ij}$)
as well as the thermo-diffusion tensor ($D_{ij}^{\,(T)} = \rho \,\phi (1-\phi)
 \,
d_{ij}^{\,(T)}$). The ratios $d_{*}^{\,(T)}/d_{*}$ (with $*\in
\{\perp,\parallel\}$) and $d_{*}^{\,(T)}/\kappa_{*}$ are called
the Soret and the Dufour coefficients, respectively (the latter
being neglected usually in liquids). The viscosity term in
(\ref{gdiss}) has the same form as in a 1-fluid description. For a
more general treatment of viscosity-like contributions cf.
Appendix.

In ad-hoc treatments of 2-fluid systems the mutual friction of the two species
 is
introduced via an interaction force $\bm{f}_{12}$ in the momentum equations
for the
single fluids, $\rho_1 \dot {\bm{v}}_{1} = {\bm{f}}_{12}$ and $\rho_2 \dot
{\bm{v}}_{2} =
-{\bm{f}}_{12}$ preserving total momentum. The force is related to the
velocity
difference, ${\bm{f}}_{12} = - \xi \,\rho_1 \,\rho_2 \,\bm{w}$ and is non-zero
only if both
fluids are present. This translates directly into $\dot {\bm{w}}
= - \xi\, \rho \,\bm{w}$
and can be compared to (\ref{wdiss}). First, in a nematic environment the
force
${\bm{f}}_{12}$ is not necessarily parallel to $\bm{w}$ due to the
possible anisotropy, rendering the $\xi$ to be a tensor $\xi_{ij} = \xi_\perp
\delta_{ij}^\perp + \xi_\parallel n_i n_j$. Then comparison with (\ref{wdiss})
gives $\xi_{ij} = \phi (1-\phi)
\xi_{ij}^\prime$, which shows that the ad-hoc choice for ${\bm{f}}_{12}$ is
the
only possible one within linear irreversible thermodynamics. Of course, there
is room for suitable nonlinear extensions (e.g. $\sim \bm{f}^3_{12}$ or $\xi$
being a function of scalar state variables like $T$, $\phi$, or $\rho$ etc.).

The reversible part of the dynamics is either dictated by symmetries or
phenomenological. The symmetry parts have been discussed in sec.\ref{dynamics}
and are listed in eqs.(\ref{sigma}--\ref{jsigma}). The phenomenological
reversible currents {\em cannot} be derived from any potential (especially not
from any kind of Hamiltonian, despite being reversible). They are
 most easily derived by
writing down all symmetry-allowed contributions to the various currents and
then
make sure that the entropy production (\ref{Rcond}) is zero. We find
\begin{eqnarray}
 \label{nrev}
Y_i^{\,(rev)} &=& -\lambda_{ijk} \nabla_j v_k -\lambda^{\,(m)}_{ijk} \nabla_j
m_k +
\beta_1 m_j \nabla_j n_i \\ \label{grev}
\sigma_{ij}^{\,(rev)} &=& -\lambda_{kji}\, \bar h_k + 2\beta_2 \,m_i \,w_j +
\beta_{2}^{\,\prime} (m_{i} g_{j} + m_{j} g_{i})\\
X_i^{\,(rev)} &=& \beta_{ij} \,\nabla_j T + \gamma_{ij} \, \nabla_j \Pi -
\nabla_j
(\lambda^{\,(m)}_{kji} \,\bar h_k) -
\beta_1 \,
\bar h_j \,\nabla_i n_j  \nonumber \\ &&+ (\beta_2 \,w_j +
\beta_{2}^{\,\prime}\,
g_{j})(\nabla_j v_i +
\nabla_{i} v_{j}) + \beta_{3} m_{j} (\nabla_{j} w_{i} - \nabla_{i} w_{j})
\nonumber
\\ && + \beta_{4} w_{j} (\nabla_{j} v_{i} - \nabla_{i} v_{j}) + \beta_{5}
(m_{i} m_{j} \nabla_{j} - \bm{m}^{2} \nabla_{i}) F
\label{wrev}
\\ \label{sigmarev}
j_i^{\,(\sigma,rev)} &=& \beta_{ij} \,m_j  \\ \label{phirev}
j_i^{\,(1,rev)} &=& \gamma_{ij} \,m_j
\end{eqnarray}
with $2 \lambda_{ijk} = \lambda_{1} \delta_{ij}^\perp \,n_k + \lambda_{2}
\delta_{ik}^\perp
\, n_j$,
$2 \lambda^{\,(m)}_{ijk} = \lambda_1^{\,(m)} \delta_{ij}^\perp \,n_k +
\lambda_2^{\,(m)}
\delta_{ik}^\perp \,n_j$,
and $\beta_{ij} = \beta_\perp \delta_{ij}^\perp + \beta_\parallel \,n_i n_j$
and $\gamma_{ij} = \gamma_\perp \delta_{ij}^\perp + \gamma_{\,\parallel} \,n_i
n_j$
and $F$ any function of the scalar variables or conjugates (e.g. $T$, $\rho$,
$\phi$). Since the term involving $F$ is already of cubic order, we will
neglect
it in the following and suppress similar terms in other equations.
Of the four flow alignment parameters $\lambda$, only three are independent as
will be discussed below. The $\beta_{ij}$-tensor in (\ref{wrev},\ref{sigmarev})
describes a reversible entropy (energy) current due to a non-zero velocity
difference as well as a change in the velocity difference due to a temperature
gradient. In the limit of large $\xi$ the $\gamma$ and $\beta$ parameters are
related to diffusion and thermodiffusion (see below).
The 1-fluid description is obtained in the limit of very large mutual
friction, $\xi\to \infty$, which implies $\bm{w} \to 0$. The usual
nematodynamics
(with an additional concentration variable) is regained, while
$\rho \xi \bm{w}$
stays finite accommodating Eqs.(\ref{wdot},\ref{wdiss},\ref{wrev}) and is
slaved by the other variables.
\\[0.5cm]

\section{Convective Velocities, Stress Division, and Concentration Dynamics
\label{stressdiv}}

In (\ref{nrev}--\ref{phirev}) we have introduced terms, which are compatible
with symmetries and $R=0$, involving quadratic nonlinearities in the
different velocities. Among them
the $\beta_1$ term has a form quite similar to the convective term in
(\ref{ndot}). Thus the actual velocity, with which $n_i$ is convected, is
$\bm{v}_{conv} = \rho^{-1} \rho_1 (\beta_1 \rho_2 +1) \bm{v}_1 + \rho^{-1}
\rho_2 (-\beta_1 \rho_1 + 1) \bm{v}_2$ and can be either $\bm{v}_1$,
$\bm{v}_2$
or something in-between, depending on $\beta_1$. Since it is hard to imagine
that $n_{i}$ is convected with a velocity larger than
$max(\mid\!\bm{v}_{1}\!\mid,\,\mid\!\bm{v}_{2}\!\mid)$, $\beta_1$ is bounded
$\rho_{1}^{-1} <
\beta_{1} < -\rho_{2}^{-1}$.
If one accepts the reasonable assumption that $n_{i}$ is convected with the
velocity of the nematic fluid $\bm{v}_{2}$, then $\beta_{1} = -\rho_{2}^{-1}$
is
fixed
(while for $\beta_{1} =0$, there is $\bm{v}_{conv} = \bm{v}$).

The choice of $\bm{v}_{conv}$
has additional implications for the flow alignment parameters $\lambda$. Since
the director does not rotate in a frame that corotates with it, the
quasicurrent $Y_{i}$ couples to the vorticity by $Y_{1}^{rot}= \epsilon_{ijk}
n_{j} \omega_{k}^{\,conv}$, where $2\omega_{k}^{\,conv}\equiv {\rm curl}
\,\bm{v}_{conv}$.
For $\bm{v}_{conv} = \bm{v}$ this implies $\lambda_{1}^{(m)}=
\lambda_{2}^{(m)}$
and $\lambda_{2}- \lambda_{1} =2$, or in
the usual parameterization $\lambda_{1}= \lambda -1$ and $\lambda_{2}= \lambda
+1$. For $\bm{v}_{conv} = \bm{v}_{2}$ the conditions are
\begin{eqnarray}
 \label{lambdacond}
 \lambda_{1}= \lambda -1, \quad \quad \lambda_{2}= \lambda+1, \quad \quad
-\rho_2 \lambda_1^{(m)} = \bar \lambda -1, \quad  \quad
-\rho_2 \lambda_2^{(m)} = \bar \lambda +1
\end{eqnarray}
In any case only two of the $\lambda$'s are independent.

The phenomenological contribution $\sim \gamma_{ij}$ in (\ref{phirev}) affects
the convection of the densities $\rho_{1}$ and $\rho_{2}$ in (\ref{contin1},
\ref{contin2}). For $\gamma_{\perp}= \gamma_{\,\parallel}=0$ the densities are
convected with $\bm{v}_{1}$ and $\bm{v}_{2}$, respectively, while $\phi$ in
(\ref{concdot}) moves with the velocity $(1/\rho)(\rho_{2} \bm{v}_{1}+
\rho_{1} \bm{v}_{2})$. All these quantities ($\rho_{1}$, $\rho_{2}$, $\rho$,
and $\phi$) are convected with $\bm{v}$ for $\gamma_{\perp}=
\gamma_{\,\parallel}=-1$.
Of course, other choices of the $\gamma$'s (made either by a theorist or by
nature!) will lead to different convection velocities.

Somewhat more involved is the question of the convective velocity for the
different momenta. Because of $\bm{g}=\rho \bm{v}$, the total momentum has to
be convected with $\bm{v}$ requiring $\beta_{2}^{\prime}=0$. If also $\bm{w}$
is convected with $\bm{v}$ then $\beta_{2}= \beta_{3}=
\beta_{4}=0$, additionally, with the consequence that also the individual
momenta, $\bm{g}_{1}$ and $\bm{g}_{2}$, are convected with $\bm{v}$. On the
other hand, for $\bm{g}_{1}$ and $\bm{g}_{2}$ to be convected with
$\bm{v}_{1}$ and $\bm{v}_{2}$, respectively, i.e.
\begin{eqnarray}
 \label{g1dot}
\dot g_{i}^{\,(1)} + \nabla_{j} g_{i}^{\,(1)} v_{j}^{\,(1)} + X_{i}^{\,(1)}
&=& 0 \\
\dot g_{i}^{\,(2)} + \nabla_{j} g_{i}^{\,(2)} v_{j}^{\,(2)} + X_{i}^{\,(2)} &=&
0
\label{g2dot}
\end{eqnarray}
where
\begin{eqnarray}
 \label{X1allg}
X_{i}^{\,(1)} &=& \,\,\,\frac{\rho_{1}\rho_{2}}{\rho} X_{i} + \frac{\rho_{1}}
{\rho}
\nabla_{j} \sigma_{ij}-  \frac{\rho_{1}}{\rho}w_{i} \nabla_{j} m_{j} -
m_{j}\nabla_{j} v_{i}^{\,(1)} \\
X_{i}^{\,(2)} &=& - \frac{\rho_{1}\rho_{2}}{\rho} X_{i} + \frac{\rho_{2}}{\rho}
\nabla_{j} \sigma_{ij}-  \frac{\rho_{2}}{\rho}w_{i} \nabla_{j} m_{j} +
m_{j}\nabla_{j} v_{i}^{\,(2)}
\label{X2allg}
\end{eqnarray}
in order to be compatible with (\ref{gdot},\ref{wdot}), the $\beta$-parameters
have to be $\beta_{2}= 1/2$, $\beta_{3}= (1/\rho_{1})- (1/\rho_{2})$, and
$\beta_{4}=1/2$
, thus ensuring that $X_{i}$ and $X_{2}$ do not contain additional transport
terms. This choice of parameters results in $(\rho_{2} \bm{v}_{1} + \rho_{1}
\bm{v}_{2})/\rho$ to be the convective velocity for $\bm{w}$ (which is the
same as for $\phi$, when $\rho_{1,2}$ are convected with $\bm{v}_{1,2}$). In
addition the momentum current density due to flow then reads $g_{j}v_{i} +
w_{j} m_{i}= \rho_{1} v_{i}^{\,(1)}v_{j}^{\,(1)} + \rho_{2} v_{i}^{\,(2)}
v_{j}^{\,(2)}$, which is the expected expression.

The terms proportional to $\bar h_i$ in (\ref{grev},\ref{wrev})
constitute forces due to the
nematic orientational elasticity. Generally they act on both fluids. Using
(\ref{X1allg},\ref{X2allg}) they read in linearized form
\begin{eqnarray}
 \label{f1}
 \dot g_i^{\,(1)}\!\mid_{nem} &=& \phi
\,(\lambda_{kji} + \rho_2 \lambda^{\,(m)}_{kji})\,
\nabla_j \bar h_k \\
 \dot g_i^{\,(2)}\!\mid_{nem} &=& (1-\phi)\,
(\lambda_{kji} - \rho_1 \lambda^{\,(m)}_{kji})\,
\nabla_j \bar h_k \label{f2}
\end{eqnarray}
Hence, for $\lambda_{ijk}^{\,(m)}=0$
($\lambda_{1}^{\,(m)}=0=\lambda_{2}^{\,(m)}$) this nematic force is
distributed on fluid 1 and fluid 2 according to the ratio of
$\rho_{1}/\rho_{2}$. It should be noted, however, that this kind of nematic
stress division is only compatible with the choice of $\bm{n}$ being convected
by $\bm{v}$, while it is incompatible with the choice of $\bm{v}_2$ as the
convective velocity for $\bm{n}$ (cf. (\ref{lambdacond})).
Another reasonable case for the stress division problem is obtained for
$\lambda_{ijk} = - \rho_{2} \lambda_{ijk}^{\,(m)}$
($\lambda_{1}=-\rho_{2}\lambda_{1}^{\,(m)}$ and
$\lambda_{2}=-\rho_{2}\lambda_{2}^{\,(m)}$). In that case the nematic force
only acts on fluid 2 (the nematic component). This case is compatible with
fluid 2 being convected
with $\bm{v}_{2}$ and the force then
reads
\begin{eqnarray}
 \dot g_i^{\,(2)}\!\mid_{nem} &=&\frac{1}{2}\,
(( \lambda_{} - 1) \delta_{kj}^{\perp} n_{i} + ( \lambda +1)
\delta_{ki}^{\perp} n_{j}) \nabla_{j} \bar h_{k} =
\lambda_{kji}
\nabla_{j} \bar h_{k}
\label{f3}
\end{eqnarray}
which is the form familiar from 1-fluid nematics. Thus,
 the so-called stress division problem (how $\bar h_i$ in
the stress tensor is divided between the two fluids) depends not only on
specific material properties expressed by the phenomenological parameters
$\lambda^{\,(m)}_{1,2}$ and $\lambda_{1,2}$, but is also intrinsically linked
to the question of the appropriate convection velocity.

In order to verify experimentally the choices above, it seems to be difficult
to directly measure specific convection velocities or the division of the
nematic stress. However, there are situations, where these choices can be
verified indirectly. Linearizing the dynamic equation
for the relative velocity (\ref{wdot},\ref{X},\ref{wdiss},\ref{wrev}) and
Fourier transform it w.r.t. time, $\bm{w}$ can be expressed by all the other
variables.
This can be used to eliminate $\bm{w}$ e.g. from the linearized dynamic
equation for the concentration (\ref{concdot}) leading to
\begin{equation}
\label{concdotred}
i \omega \phi -  d_{ij}^{\,eff} \nabla_i \nabla_j \Pi -
\frac{\rho_1 \rho_2}{\rho^2} d_{ij}^{\,(T)\,eff}
\nabla_i \nabla_j T + \lambda^{(\phi)} (\bm{n}\cdot\bm{\nabla}) div \bm{h}=0
\end{equation}
where contributions of order $O(\nabla^4)$ have been neglected. The effective
diffusion and thermo-diffusion (Soret) coefficients have
got additional frequency dependent contributions due to the 2-fluid degree of
freedom
\begin{eqnarray}
\label{deff}
d_*^{\,eff} &=& d_* + \frac{\rho_1 \rho_2}{\rho^2}\, \frac{(1+\gamma_*)^2}
{\rho \,\xi_* + i \omega} \\ \label{dTeff}
d_*^{\,(T)\,eff} &=& d_*^{\,(T)} + \frac{\beta_* (1+ \gamma_*)}{\rho \,\xi_*
+ i \omega}
\end{eqnarray}
where the subscript $_*$ stands for either $_{\parallel}$ or $_{\perp}$.
There is also a dynamic coupling to the nematic degree of freedom due to
\begin{equation}
\label{lambdaphi}
\lambda^{(\phi)} = \frac{\rho_1 \rho_2}{2 \rho^2}\Bigl(\lambda_1^{(m)}
\frac{1+ \gamma_{\parallel}}{\rho \xi_{\parallel} + i \omega} + \lambda_2^{(m)}
\frac{1+ \gamma_{\perp}}{\rho \xi_{\perp} + i \omega} \Bigr)
\end{equation}
These possible additions to the concentration dynamics, however, depend on the
choices for the convection velocities as well as on the way how the nematic
stress has been divided among the two fluids. Assuming the densities
$\rho_{1,2}$ to be convected with the mean velocity $\bm{v}$ (implying
$\gamma_{\parallel} = \gamma_{\perp}= -1$) the additional contributions to
diffusion and thermo-diffusion are all zero as well as the coupling to the
nematic director. On the other hand, for $\rho_{1,2}$ to be convected with
$\bm{v}_{1,2}$ respectively ($\gamma_{\parallel} = \gamma_{\perp}= 0$), both,
diffusion and thermo-diffusion show a
dispersion step around $\omega \approx \rho \,\xi$. For a nematic stress
division among fluid 1 and 2 according to the ratio $\rho_1/\rho_2$ (implying
$\lambda_1^{(m)} = \lambda_2^{(m)} =0)$ there is no dynamic influence of the
nematic degree of freedom on the concentration ($\lambda^{(\phi)}=0$), while
for
any other choice there is one. In particular, if only fluid 2 carries
nematic stress (and $\rho_{1,2}$ are convected with $\bm{v}_{1,2}$,
respectively), this dynamic coupling is given by $\lambda^{(\phi)} = -
(\rho_1/2\rho^3) [\bar \lambda (\xi_{\parallel}^{-1}+ \xi_{\perp}^{-1}) -
\xi_{\parallel}^{-1}+ \xi_{\perp}^{-1}]$ for strong friction ($\rho \xi_* \gg
\omega$).
\\[0.5cm]

\section{Simplified 2-Fluid Nematic Equations \label{simplified}}

In the preceding sections we have derived the most general and complete set
of 2-fluid equations for a nematic and Newtonian mixture. Special emphasis has
been laid on the correct form of the nonlinearities that come with the 2-fluid
description. However, these equations are for most purposes unnecessarily
complicated and can be simplified using reasonable assumptions. Starting from
the correct general equations such assumptions, clearly spelled out, lead to
controlled approximations and to a set of 2-fluid equations, whose limitations
and implicit assumptions are clear and well defined in contrast to most ad-hoc
approaches.

Here we want to display explicitly 2-fluid hydrodynamics for a nematic/simple
fluid mixture under the following assumptions, \\ a)
convection with natural velocities (for $\bm{n}$, $\bm{g}_{2}$, $\rho_{2}$ and
$\bm{g}_{1}$, $\rho_{1}$ this is $\bm{v}_{2}$ and $\bm{v}_{1}$, respectively,
or
explicitly $\beta_{1} = -\rho_{2}^{-1}$,
$\beta_{2} = \beta_{4} = 1/2$, $\beta_{3}= \rho_{1}^{-1} - \rho_{2}^{-1}$,
$\gamma_{\perp} = 0 = \gamma_{\,\parallel}$ and eq.(\ref{lambdacond})); \\
 b) the
linearized orientation-elastic force acts on the nematic fluid (index 2) only
(i.e.  $\lambda_{1,2} = - \rho_2 \lambda_{1,2}^{\,(m)}$); \\
c) global
incompressibility, $\delta \rho =0$ (i.e. $\delta \rho_{1} = - \delta
\rho_{2}$);\\
d) neglecting the phenomenological reactive entropy current ($\beta_{ij} =
0$);\\ e) linearizing the phenomenological dissipative currents, but keeping
quadratic nonlinearities otherwise.

Then
the following set of equations is obtained:\\ The incompressibility
condition
\begin{eqnarray}
 \label{sinc}
0 &=& {\rm div}\, {\bm{v}} \\
\label{sinc2}
{\rm or}\quad\quad 0 &=& \bm{w}\cdot \bm{\nabla} \rho_1 + \rho_{1} {\rm div} \,
\bm{v}_{1} + \rho_{2} {\rm div} \,\bm{v}_{2}  \\
\label{sinc3}
{\rm or}\quad\quad 0 &=& \bm{w}\cdot \bm{\nabla} \phi + \phi \,{\rm div}(1-
\phi)\bm{w}
 - (1-\phi) \,{\rm div} \phi \bm{w}
\end{eqnarray}
the concentration dynamics
\begin{eqnarray}
 \label{sconcdyn}
\dot \phi +  \nabla_{i} \left( \phi v_i +  \phi (1-\phi) w_{i}\right) -
   d_{ij} \nabla_i \nabla_{j}(\mu_1 - \bar \mu_2) - \phi(1-\phi)
   d_{ij}^{\,(T)} \nabla_{j}\nabla_i T
 &=& 0  \\
\label{sconcdyn2}  {\rm or} \quad\quad
\dot \rho_{1} + \bm{v}_{1}\cdot \bm{\nabla} \rho_{1} + \rho_{1} {\rm div}
\bm{v}_{1}
- \rho \, d_{ij} \nabla_i  \nabla_j (\mu_1 - \bar \mu_2) - \frac{\rho_1\rho_2}
{\rho}  d_{ij}^{\,(T)} \nabla_i
 \nabla_j T &=& 0 \\
\label{sconcdyn3}  {\rm or} \quad\quad
\dot \rho_{2} + \bm{v}_{2}\cdot \bm{\nabla} \rho_{2} + \rho_{2} {\rm div}
\bm{v}_{2}
+ \rho \, d_{ij} \nabla_i \nabla_j (\mu_1 - \bar \mu_2) + \frac{\rho_1\rho_2}
{\rho} d_{ij}^{\,(T)}
\nabla_i\nabla_j T &=& 0
\end{eqnarray}
the entropy dynamics (heat conduction equation)
\begin{equation}
\label{ssigmadyn}
\dot \sigma + v_{i} \nabla_{i} \sigma -  \kappa_{ij} \nabla_{i} \nabla_{j} T
-
\frac{\rho_1\rho_2}{\rho} d_{ij}^{\,(T)} \nabla_i \nabla_{j} (\mu_1 - \bar
\mu_2)
= 0
\end{equation}
the nematic director dynamics
\begin{equation}
\label{sndyn}
\dot n_{i} + v_{j}^{\,(2)} \nabla_{j} n_{i}
- \lambda_{ijk}\nabla_{j} v_{k}^{\,(2)} -\frac{\rho_1}{\rho_2}
\lambda_{ijk} w_k \nabla_j \phi
+\gamma_{1}^{{-1}} \delta_{ij}^{\perp}
\,\bar h_{j} =0.
\end{equation}
There is a (nonlinear) coupling to the concentration variable, which is not
possible in the 1-fluid description.
\\  For
the momentum balance of the two different species we get
\begin{eqnarray}
 \label{sg1dyn}
&&\!\!\!\!\!\!\!\!\!\!\!\!\!\! \rho_{1} \dot v_{i}^{\,(1)} + \rho_{1}
v_{j}^{\,(1)} \nabla_{j} v_{i}^{\,(1)}
+ \frac{\rho_1}{\rho} \nabla_i (p +\frac{1}{2} \rho_{2} (\bm{v}_{1}^{2} -
\bm{v}_{2}^{2})) + \frac{\rho_1 \rho_2}{\rho} \nabla_i
(\mu_1 - \bar \mu_2) + \frac{\rho_1}{\rho} \nabla_j (\Psi_{kj} \nabla_i n_k)
\nonumber \\ &+&\!\!\!\!  \frac{\rho_1}{\rho} \bar h_{j} \nabla_{i} n_{j}
+ \frac{\rho_1}{\rho_2}
\lambda_{kji} \bar h_k \nabla_j \phi
 +  \xi_{ij} \rho_1 \rho_2 w_j - \nu_{ijkl}^{\,(1)}
\nabla_j \nabla_l v_k^{\,(1)}  = 0
\\
&&\!\!\!\!\!\!\!\!\!\!\!\!\!\! \rho_{2} \dot v_{i}^{\,(2)} + \rho_{2}
v_{j}^{\,(2)} \nabla_{j} v_{i}^{\,(2)}
+ \frac{\rho_2}{\rho} \nabla_i (p - \frac{1}{2} \rho_{1} (\bm{v}_{1}^{2}-
\bm{v}_{2}^{2})) - \frac{\rho_1 \rho_2}{\rho} \nabla_i
(\mu_1 - \bar \mu_2) +  \frac{\rho_2}{\rho} \nabla_j (\Psi_{kj} \nabla_i n_k)
\nonumber  \\  &-&\!\!\!\! \frac{\rho_1}{\rho} \bar h_{j} \nabla_{i} n_{j} -
 \frac{\rho_1}{\rho_2}
\lambda_{kji} \bar h_k \nabla_j \phi -
\nabla_j (\lambda_{kji} \bar h_k)  -
 \xi_{ij} \rho_1 \rho_2 w_j - \nu_{ijkl}^{\,(2)}
\nabla_j \nabla_l v_k^{\,(2)} = 0 \label{sg2dyn}
\end{eqnarray}
Note that although we made the approximation that the linear
orientational-elastic stress does only act on fluid 2, there are
inevitably nonlinear contributions to fluid 1, too. There is also
a (nonlinear) coupling of fluid 1 to the concentration, if nematic
distortions ($\bar h_{i} \neq 0$) are present. In
(\ref{sg1dyn},\ref{sg2dyn}) cross-viscosities have been neglected
(cf. Appendix).

In order to facilitate actual calculations we also give
eqs.(\ref{sg1dyn},\ref{sg2dyn}) as dynamic equations for the total momentum
and for the relative velocity
\begin{eqnarray}\label{sgdyn}
\!\!\rho \dot v_{i} + \nabla_{i}p + \rho \nabla_{j}\Bigl( v_{i}v_{j} +
\phi(1-\phi) w_{i}w_{j} \Bigr) + \nabla_{j}\Bigl(
\Psi_{kj}\nabla_{i}n_{k} - \lambda_{kji} \bar
h_{k} \Bigr) \nonumber \quad\quad\quad\quad&& \\
- \nu_{ijkl} \nabla_{j}\nabla_{l}v_{k} &\!\!=\!\!& 0 \quad
\\ \label{swdyn} \dot w_{i} + \Bigl( v_{j} + (1-2\phi) w_{j}
\Bigr) \nabla_{j}w_{i} + \nabla_{i}\Bigl(\mu_{1} - \bar \mu_{2} +
\bm{v \cdot w} + (\frac{1}{2}-\phi) \bm{w}^{2} \Bigr) + \rho
\xi_{ij} w_{j}  && \nonumber
\\
+ \frac{1}{\rho_{2}}\bar h_{j}\nabla_{i}n_{j} +
\nabla_{j}\Bigl(\frac{1}{\rho_{2}} \lambda_{kji} \bar h_{k} \Bigr)
&\!\! =\!\!& 0
\end{eqnarray}
Note that the approximation for the viscosities made in
(\ref{sgdyn},\ref{swdyn}) is not compatible with that used in
(\ref{sg1dyn},\ref{sg2dyn}); their
interrelation is discussed in the Appendix.

Due to the incompressibility condition  the pressure is no
longer an independent
variable nor is it given by the other
variables (i.e. (\ref{Duhem}) or (\ref{Duhem2}) cannot be used), but it
serves as an auxiliary quantity to ensure the
incompressibility for all times, i.e. ${\rm div} \dot{\bm{v}} =0$, which leads
 to
the very complicated condition that determines $\delta p$
\begin{eqnarray}
\label{laplacep}
\Delta p &=& -\nabla_{i} \nabla_j \,(\rho_{1} v_{i}^{(1)} v_{j}^{(1)} +
\rho_{2} v_{i}^{(2)} v_{j}^{(2)} ) - \nabla_i \nabla_j\, (\Psi_{kj} \nabla_i
n_k)\\ \nonumber&&
 +   \lambda  \nabla_i \nabla_{j}\, (\delta_{kj}^{\perp} n_{i} \bar h_{k})
+  \nu_{ijkl}  \nabla_i \nabla_j \nabla_l  v_k
\end{eqnarray}
Although $\delta p$ does not show up in the dynamical equations, it is still
present in boundary conditions etc. and it contains combinations of the
viscosities different from those present in the incompressible dynamical
equations.\footnote{ Incompressibility 'reduces' the number of independent
components of the viscosity tensor from five to three (in the uniaxial case), only if
a redefinition of the pressure is done, cf. \cite{mbuch}
p.41f and \cite{leslie}}
In contrast to 1-fluid descriptions for simple fluids, where the
incompressibility condition leads to a considerable mathematical
simplification, this is no longer the case for a 2-fluid description due the
complicated form of (\ref{laplacep}), even if incompressibility is a very good
approximation in physical terms.

Of the statics (\ref{susT}--\ref{susmu}) only the following equations remain
\begin{eqnarray}
 \label{ssusT}
\delta T &=& T C_{V}^{{-1}} \, \delta \sigma + \alpha_{\phi}^{-1} \delta \phi
\\ \label{ssusmu}
\delta (\mu_{1} - \bar \mu_{2}) &=& \rho^{-1} \kappa_{\phi}^{-1} \, \delta
\phi + \rho^{-1} \alpha_{\phi}^{-1} \, \delta \sigma
\end{eqnarray}
with $\delta \phi = \rho^{-1} \delta \rho_{1} = - \rho^{-1} \delta \rho_{2}$,
while
(\ref{Frank}) and (\ref{Franknl}) are unchanged.
Note that $\delta \mu$ is not needed, but follows from $\delta p$ via eq.
(\ref{Duhem2}).
\\[0.5cm]

\section{Isotropic Viscoelastic Fluids \label{polymers}}

In this section we discuss the 2-fluid description of isotropic viscoelastic
fluids
by choosing a Newtonian fluid as fluid 1 and an elastic medium as fluid 2. The
latter can be a permanent network (showing e.g. diffusion) or a temporary one
relaxing on a finite time scale. The considerations for setting up a complete
nonlinear 2-fluid description for such systems is quite similar to that of the
2-fluid nematic discussed in detail in the previous sections - only that the
nematic
degree of freedom $\nabla_j n_i$ has to be replaced by the Eulerian strain
tensor $U_{ij}$, which we use to describe the elastic degree of freedom. In
the following we present an abbreviated discussion, starting with the general
energy expression.

\subsection{Thermodynamics}

In analogy with the development in section \ref{princ}, the general energy
expression for an isotropic elastomer network immersed in a Newtonian solvent is
given by
\begin{equation}
 \label{penergy}
E= \epsilon \,V = \int \epsilon\, dV = E(M_1,\, M_2,\, V,\, \bm{G_1},\,
\bm{G_2},
\, S,\, M_2 U_{ij})
\end{equation}
from which the conjugate quantities are derived.
The elastic stress, conjugate to the strain, is $\Phi_{ij}^{\prime} = \partial
 E / \partial
(M_2 U_{ij}) =
\partial \epsilon /
\partial (\rho_2 U_{ij}) \equiv \rho_{2}^{-1} \Phi_{ij}$, while the definitions
of the other
conjugates (\ref{partial}) remain unchanged (except that they are to be taken
at constant $U_{ij}$ rather than constant $\nabla_{j}n_{i}$). With the new
definition $\bar
\mu_{2} = \mu_{2} + \rho^{-1} \Phi_{ij} U_{ij}$ the expressions for the
pressure (\ref{pressure},\ref{pressure2}) and the relations of the different
sets of conjugates
(\ref{conjugates2}) remain unchanged, while the Gibbs and Gibbs-Duhem
relations read
\begin{eqnarray}
 \label{pGibbs}
d\epsilon &=& T d\sigma + \mu_1 \,d\rho_1 + \bar \mu_2 \,d\rho_2 + \bm{v}_1
\cdot
d\bm{g}_1 + \bm{v}_2 \cdot d\bm{g}_2  + \Phi_{ij} \,d U_{ij}
\\ \label{pGibbs2}
 &=& T\, d\sigma + \Pi\, d\phi +  \mu\, d\rho + \bm{v} \cdot
d\bm{g}  + \bm{m} \cdot d\bm{w} + \Phi_{ij} \,d U_{ij} \\
\label{pDuhem}
dp &=& \sigma \,dT + \rho_1 \,d\mu_1 + \rho_2 \,d\bar \mu_2
+ \bm{g}_1 \cdot d\bm{v}_1 + \bm{g}_2 \cdot d\bm{v}_2 - \Phi_{ij} \,d U_{ij}\\
\label{pDuhem2}
&=& \sigma\, dT + \rho\, d\mu
+ \bm{g} \cdot d\bm{v} - \bm{m} \cdot d\bm{w} - \Phi_{ij} \,d U_{ij}
\end{eqnarray}
Rotational invariance of the Gibbs relation (\ref{pGibbs},\ref{pGibbs2}) leads to
the condition
\begin{equation}
\label{protcond}
U_{ik} \Phi_{kj} = U_{jk} \Phi_{ki}
\end{equation}
which, as is seen later on, ensures the stress tensor to be symmetric.

\subsection{Statics}

The conjugate quantities defined by the Gibbs relation
(\ref{pGibbs},\ref{pGibbs2}) are linked to the variables by a set
of phenomenological equations containing static susceptibilities
as parameters. This constitutes the static part of the
hydrodynamics. Instead of the nematic molecular fields $h_{i}$ and
$\psi_{ij}$ we now have the elastic stress $\Phi_{ij}$ as
conjugate field. As a symmetric 2-rank tensor it consists of a
scalar quantity, the trace $\Phi_{ii}$ and the deviator
$\Phi_{ij}^{\,(0)} = \Phi_{ij} -(1/3)\delta_{ij}\Phi_{kk}$. Being
a scalar $\Phi_{kk}$ can couple to the other scalar variables like
densities, concentration or entropy, just like the 3 other scalar
conjugates \{$T,\,\Pi,\, \mu$ or $\mu_{1},\, \mu_{2}$\} by
\begin{eqnarray}
\label{psusT}
\delta T &=& \frac{T}{C_{V}}\delta \sigma + \frac{1}{\rho \alpha_{1}}
\delta \rho_{1} + \frac{1}{\rho \alpha_{2}}
   \delta \rho_{2}  + \frac{1}{\alpha_{3}} U_{kk} \\ \nonumber
  &=&   \frac{T}{C_{V}}\delta \sigma +  \frac{1}{\alpha_{\phi}} \delta
\phi + \frac{1}{\rho \alpha_{\rho}}  \delta \rho + \frac{1}{\alpha_{3}} U_{kk}
 \\
\label{psusmu1}
\mu_{1} &=& \frac{1}{\rho^{2} \kappa_{1}} \delta \rho_{1}+
\frac{1}{\rho^{2} \kappa_{3}} \delta \rho_{2} +
\frac{1}{\rho \alpha_{1}} \delta \sigma + \frac{1}{\rho \kappa_{4}} U_{kk}\\
\label{psuspi}
\Pi &=& \frac{1}{\kappa_{\phi}} \delta \phi +
\frac{1}{\rho \kappa_{\pi}} \delta \rho +
\frac{1}{\alpha_{\phi}} \delta \sigma  + \frac{1}{\kappa_{u}}U_{kk} +
\bm{w}\cdot \bm{g} +\rho
  \bm{w}^{2}(1-2\phi) \\
\label{psusmu2}
\bar \mu_{2} &=& \frac{1}{\rho^{2} \kappa_{2}} \delta \rho_{2}+
\frac{1}{\rho^{2} \kappa_{3}} \delta \rho_{1} +
\frac{1}{\rho \alpha_{2}} \delta \sigma + \frac{1}{\rho \kappa_{5}} U_{kk}\\
\label{psusmu}
\mu &=& \frac{1}{\rho^{2} \kappa_{\mu}} \delta \rho +
\frac{1}{\rho \kappa_{\pi}} \delta \phi +
\frac{1}{\rho \alpha_{\rho}} \delta \sigma + \frac{1}{\rho \kappa_{\rho}} U_{kk}
+ \bm{w}^{2} \phi (1-\phi) \\
\label{Phitrace}
\Phi_{kk} &=& c_{l} U_{kk} + \frac{1}{\alpha_{3}} \delta \sigma + \frac{1}{\rho
  \kappa_{4}} \delta \rho_{1} + \frac{1}{\rho
  \kappa_{5}} \delta  \rho_{2}\\ \nonumber
&=&  c_{l} U_{kk} + \frac{1}{\alpha_{3}} \delta \sigma + \frac{1}{\rho
  \kappa_{u}} \delta \phi + \frac{1}{\rho
  \kappa_{\rho}} \delta  \rho \\
\label{Phidev}
\Phi_{ij}^{\,(0)}&=& c_{tr} (U_{ij} - \frac{1}{3}\delta_{ij}U_{kk})
\end{eqnarray}
where -- in addition to (\ref{alpha1}--\ref{alpha2})
\begin{eqnarray} \label{kappa1}
{\kappa_{\rho}}^{-1} &=& \phi {\kappa_{4}}^{-1} + (1-\phi)
{\kappa_{5}}^{-1} \\
{\kappa_{u}}^{-1} &=& {\kappa_{4}}^{-1} -
{\kappa_{5}}^{-1} \label{kappa2}
\end{eqnarray}
involving 2 new generalized compressibilities $\kappa_{4,5}$ or
$\kappa_{u,\rho}$ and one expansion
coefficient $\alpha_{3}$ related to the trace of the elastic strain $U_{kk}$.
It should be noted that for real solids at finite temperatures $U_{kk} \neq
\delta \rho / \rho$ in contrast to ideal elasticity theory. The reason are the
point defects, which allow not only the dissipative motion described above,
but also static temperature and pressure changes due to $U_{kk}$ even at
constant density. The new static susceptibilities $c_{l}$ and $c_{tr}$ are the
usual elastic moduli of Hooke's law. The longitudinal one is related (in
addition to
the compressibility $\kappa_{\mu}$) to the sound velocity. The transverse
modulus leads to
transverse sound, which is however relaxing due to
(\ref{Udiss}) if $\zeta_l$ and $\zeta_{tr}$ are not zero.
\\[0.5cm]

\subsection{Dynamics}

The dynamical equations for the elastomeric and solvent degrees of freedom are
\begin{eqnarray}
\label{pcontin}
\dot \rho + \nabla_j \rho \, v_j &=& 0 \\ \label{pconcdot}
\dot \phi + v_j \nabla_j \phi +  \rho^{-1} \nabla_i  \left( \rho \phi (1-\phi)
  w_i+ j_i^{\,(1)}
\right)  &=& 0\\
\label{pepsdot}
\dot \epsilon + \nabla_j (\epsilon + p) v_j  + \nabla_i j_i^{\,(\epsilon)} &=&
 0
\\
\label{psigmadot}
\dot \sigma + \nabla_j (\sigma v_j +
j_i^{\,(\sigma,rev)} + j_i^{\,(\sigma,dis)})&=&  R/T \\
\label{pwdot}
\dot w_i + v_j \nabla_j w_i + \nabla_i  \Pi +
X_i^{\,(rev)} + X_i^{\,(dis)}  &=& 0 \\
\label{pgdot}
\dot g_i + \nabla_j g_i v_j + \nabla_{i} p +\nabla_j (-\Phi_{ij}  + \Phi_{jk}
U_{ik} + \Phi_{ik}U_{jk}+
\sigma_{ij}^{\,(rev)} +
\sigma_{ij}^{\,(dis)}) &=& 0   \\  \label{Udot}
\dot U_{ij} + v_{k} \nabla_{k} U_{ij} +U_{kj}\nabla_{i}v_{k} + U_{ki}
\nabla_{j}v_{k} - A_{ij} + Z_{ij}^{\,(rev)} +
Z_{ij}^{\,(dis)} &=& 0
\end{eqnarray}
the first 5 equations have the same form as before (but different
phenomenological currents, see below).
In the dynamic equation for the strain (\ref{Udot}) there are nonlinear
couplings to the
velocity gradient
that have the form of the so-called lower convected derivative
\cite{Temmen,pleiner} and the appropriate counter terms show up in the stress tensor
as additions to the phenomenological parts. In all dynamic equations the
convective velocity chosen is the
mean velocity $\bm{v}$, since this allows a simple and thermodynamically
consistent way of writing the equations. However, as in the case discussed
previously there are
phenomenological terms in the reversible currents that allow a different
choice of the convective velocities (see below).

\subsection{Currents}

Following the previous development, we can establish the reversible and
dissipative
currents, and their constraints.

For the phenomenological parts of the currents there is the condition
\begin{equation}
\label{pRcond}
R = -j_i^{\,(\sigma,*)} \nabla_i T + \Pi\,\, \nabla_{i} j_{i}^{(1)}-
\sigma_{ij}^{\,(*)} \nabla_j v_i + \Phi_{ij}
Z_{ij}^{\,(*)} + m_i \, X_i^{\,(*)} \geq 0
\end{equation}
with the equal sign ($>$ sign) for $* = rev$ ($*=dis$), respectively.

The dissipative parts of the currents introduced above can again be deduced
from a dissipation function that reads in bilinear approximation
\begin{eqnarray}
 \label{pdissipation}
2R &=& \kappa (\bm{\nabla} T)^{2} +  D (\bm{\nabla} \Pi)^{2} + 2D^{(T)}
(\bm{\nabla} T)\cdot(\bm{\nabla} \Pi) +  \xi^\prime \bm{m}^{2} \nonumber \\
&&  +
\zeta_{ijkl} \Phi_{ij} \Phi_{kl}+
\xi_{ijklmn} (\nabla_{i} \Phi_{jk})(\nabla_{l}\Phi_{mn}) +
\nu_{ijkl} (\nabla_j v_i)(\nabla_l v_k) \nonumber \\
&&  + \nu_{ijkl}^{\,(c)}
\left( (\nabla_j v_i)(\nabla_l m_k) + (\nabla_j m_i)(\nabla_l v_k) \right) +
\nu_{ijkl}^{\,(w)} (\nabla_j m_i)(\nabla_l m_k)
\end{eqnarray}
where all 4-rank material tensors have the form $\nu_{ijkl} =
\nu_{l} \delta_{ij} \delta_{kl} + (1/2) \nu_{tr}
(\delta_{ik}\delta_{jl} + \delta_{il}\delta_{jk} - (2/3)
\delta_{ij} \delta_{kl}))$ and $\xi_{ijklmn}$ contains 4
parameters $\xi_{1-4}$. In the $\zeta$-tensor $\zeta_{l}$ and
$\zeta_{tr}$ are describing the relaxation of elastic strains and
the $\xi_{1-4}$ give rise to vacancy diffusion as can be seen in
the following expressions
\begin{eqnarray}
\label{psigmadiss}
j_i^{\,(\sigma,dis)} &=& -(\partial R) / (\partial \nabla_i T) = -\kappa
\nabla_i T -   \rho \,\phi (1-\phi) \,d^{\,(T)} \nabla_i \Pi \\ \label{Udiss}
Z_{ij}^{\,(dis)} &=& (\partial R) / (\partial \Phi_{ij}) = \zeta_{ijkl}
\Phi_{kl} - \nabla_{k} (\xi_{kijlmn}  \nabla_{l} \Phi_{mn})
 \\ \label{pgdiss}
\sigma_{ij}^{\,(dis)} &=& - (\partial R) / (\partial \nabla_j v_i) = -
\nu_{ijkl}\,
\nabla_l v_k - \nu_{ijkl}^{\,(c)} \,\nabla_l m_k \\ \label{pwdiss}
X_i^{\,(dis)} &=& (\partial R) / (\partial m_i) = \xi^\prime \,m_i - \nabla_j
\left( \nu_{ijkl}^{\,(w)}\, \nabla_l w_k + \nu_{ijkl}^{\,(c)}\, \nabla_l v_k
\right) \\
\label{pphidiss}
j_i^{\,(1,dis)} &=& - (\partial R) / (\partial \nabla_i \Pi) = -\rho \,d
\nabla_i \Pi
-   \rho \,\phi (1-\phi) \,d^{\,(T)} \nabla_i T
\end{eqnarray}
where diffusion and thermodiffusion is written in the usual way with $D=\rho
d$ and $D^{(T)}=
\rho \phi (1-\phi) d^{(T)}$. For a permanent network that does not relax, the
relaxation parameters vanish ($\zeta_l = \zeta_{tr}=0$).
For the reversible parts of the currents we find
\begin{eqnarray}
 \label{Urev}
Z_{ij}^{\,(rev)} &=&  \lambda^{(U)}(\nabla_i m_j + \nabla_j m_i) +
\beta_{7} (U_{kj} \nabla_{i} m_{k} + U_{ki} \nabla_{j}
m_{k})
+\beta_{6} m_{k}\nabla_{k} U_{ij}
 \\ \label{pgrev}
\sigma_{ij}^{\,(rev)} &=&
2 \beta_2 \,m_i  \,w_j \\
\label{pwrev}
X_i^{\,(rev)} &=& 2\nabla_j (\lambda^{(U)} \Phi_{ij}) + \beta \,\nabla_i T
+ \gamma \nabla_i \Pi
- \beta_{6} \Phi_{kj} \nabla_{i} U_{kj}+
 \beta_2 \,w_j (\nabla_j v_i +
\nabla_{i} v_{j})  \nonumber \\&&+  \nabla_{j}\beta_{7} (\Phi_{kj}
U_{ik} + \Phi_{ki} U_{jk}) +\beta_{3} m_{j} (\nabla_{j}w_{i}- \nabla_{i}w_{j})
+ \beta_{4} w_{j}
(\nabla_{j}v_{i}- \nabla_{i}v_{j}) \quad\quad \\
\label{psigmarev}
j_i^{\,(\sigma,rev)} &=& \beta \,m_i \quad
\\
\label{pphirev}
j_i^{\,(1,rev)} &=& \gamma \,m_i
\end{eqnarray}
\subsection{\label{pconvstress} Convection, Stress, and Concentration Dynamics}

As in the case of the 2-fluid nematics the velocities which with
the variables are convected can be tuned by choosing special
values for the coefficients $\beta_{n}$ and $\gamma$. E.g. for
$\gamma=0$ the densities $\rho_{1,2}$ are convected with
$\bm{v}_{1,2}$, respectively (and the total density $\rho$ and the
concentration $\phi$ with $\bm{v}$ and $(1/\rho)(\rho_{2}
\bm{v}_{1} + \rho_{1}\bm{v}_{2})$, respectively), while for
$\gamma =-1$ all 4 quantities are convected with $\bm{v}$.
Similarly, for $\beta_{4}= 1/2 = \beta_{2}$ and $\beta_{3}=
(1/\rho_{1}) - (1/\rho_{2})$ the momenta $\bm{g_{1,2}}$ are
convected with $\bm{v}_{1,2}$ (and the total momentum $\bm{g}$ and
the relative velocity $\bm{w}$ with $\bm{v}$ and
$(1/\rho)(\rho_{2} \bm{v}_{1} + \rho_{1}\bm{v}_{2})$,
respectively). For $\beta_{6}=-1/\rho_{2}$ the strain $U_{ij}$ is
convected with $\bm{v}_{2}$ and for $\beta_{7}=-1/\rho_{2}$ the
lower convected derivative contributions in (\ref{Udot})
effectively come with $\bm{v}_{2}$ (producing an additional cubic
term in (\ref{pwrev}) $\sim \nabla_{i}\rho_{2}$, which can be
neglected as other cubic terms). Even the convection of the
entropy can be tuned by choosing $\beta\equiv \beta_{0}+
\beta_{00}\sigma$ where $\beta_{00}=1/\rho_{1}$, $=0$,
$=-1/\rho_{2}$ leads to the convective velocity to be
$\bm{v}_{1}$, $\bm{v}$, $\bm{v}_{2}$, respectively.

The distribution of the elastic stress among the two fluids is governed by the
coefficient $\lambda^{(U)}$. For, respectively, $2 \lambda^{(U)}= 1/\rho_2$, $
=-1/\rho_1$,
or $=0$, the elastic stress is carried by fluid 2, fluid 1, or is equally
distributed between them.

As in the case of 2-fluid nematics we can linearize and Fourier transform the
dynamic equations, thus eliminating $\bm{w}$ from e.g. the concentration
dynamics. Neglecting fourth order gradient terms we get
\begin{equation}
\label{pconcdoteff}
i \omega \phi -  d^{\,eff} \Delta \Pi - \frac{\rho_1 \rho_2}{\rho^2}
d^{(T)\,eff} \Delta T - 2 \lambda^{(\phi)} \nabla_i \nabla_j \Phi_{ij}=0
\end{equation}
with frequency dependent effective diffusion and thermo-diffusion coefficients
\begin{eqnarray}
\label{pdeff}
d^{\,eff} &=& d + \frac{\rho_1 \rho_2}{\rho^2}\, \frac{(\gamma +1)^2}
{\rho \,\xi + i \omega} \\ \label{pdTeff}
d^{\,(T)\,eff} &=& d^{\,(T)} + \frac{\beta (\gamma+1)}{\rho \,\xi
+ i \omega}
\end{eqnarray}
and the dynamic coupling to the elastic degree of freedom by
\begin{equation}
\label{plambdaphi}
\lambda^{(\phi)} = \frac{\rho_1 \rho_2}{\rho} \lambda^{(U)}\frac{1+
\gamma}{\rho \xi + i \omega}
\end{equation}
Again these possible additions to the concentration dynamics, however,
depend on the
choices for the convection velocities as well as on the way how the nematic
stress has been divided among the two fluids. Assuming the densities
$\rho_{1,2}$ to be convected with the mean velocity $\bm{v}$ (implying
$\gamma= -1$) the additional contributions to
diffusion and thermo-diffusion are all zero as well as the coupling to the
nematic director. On the other hand, for $\rho_{1,2}$ to be convected with
$\bm{v}_{1,2}$ respectively (e.g. $\gamma=0$), both, diffusion and
thermo-diffusion show a
dispersion step around $\omega \approx \rho \,\xi$. For the elastic stress
division among fluid 1 and 2 according to the ratio $\rho_1/\rho_2$ (implying
$\lambda_1^{(U)} =0)$ there is no dynamic influence of the
elastic degree of freedom on the concentration, while
for
any other choice there is one. In particular, if only fluid 2 carries
elastic stress ($\lambda_1^{(U)} =1/\rho_2$), this dynamic coupling is given
by $\lambda^{(\phi)} = \rho_1/\rho^2 \xi$ for strong friction ($\rho \xi \gg
\omega$).

\subsection{Simplified elastomeric two-fluid equations}

In the preceding sections we have derived the most general and complete set
of 2-fluid equations. These equations are for most purposes unnecessarily
complicated and can be simplified using reasonable assumptions. Starting from
the correct general equations such assumptions, clearly spelled out, lead to
controlled approximations and to a set of 2-fluid equations, whose limitations
and implicit assumptions are clear and well defined in contrast to most ad-hoc
approaches.

Here we want to display explicitly 2-fluid hydrodynamics under the following
assumptions, \\ a)
convection with natural velocities (for $U_{ij}$, $\bm{g}_{2}$, $\rho_{2}$ and
$\bm{g}_{1}$, $\rho_{1}$ this is $\bm{v}_{2}$ and $\bm{v}_{1}$, respectively,
or
explicitly $\beta_{7} = - \rho_{2}^{-1} = \beta_6$,
$\beta_{2} = \beta_{4} = 1/2$, $\beta_{3}= \rho_{1}^{-1} - \rho_{2}^{-1}$,
$\gamma = 0$); \\
 b) the
linearized elastic force acts on the elastomeric fluid (index 2) only
(i.e.  $2\lambda^{(U)} =\rho_{2}^{-1} $); \\
c) global
incompressibility, $\delta \rho =0$ (i.e. $\delta \rho_{1} = - \delta
\rho_{2}$);\\
d) linearizing the phenomenological dissipative currents, but keeping
quadratic nonlinearities otherwise.

Then
the following set of equations is obtained:\\ The incompressibility
condition
\begin{eqnarray}
 \label{psinc}
0 &=& {\rm div}\, {\bm{v}} \\
\label{psinc2}
{\rm or}\quad\quad 0 &=& \bm{w}\cdot \bm{\nabla} \rho_1 + \rho_{1} {\rm div} \,
\bm{v}_{1} + \rho_{2} {\rm div} \,\bm{v}_{2}  \\
\label{psinc3}
{\rm or}\quad\quad 0 &=& \bm{w}\cdot \bm{\nabla} \phi + \phi \,{\rm div}(1-
\phi)\bm{w}
 - (1-\phi) \,{\rm div} \phi \bm{w}
\end{eqnarray}
the concentration dynamics
\begin{eqnarray}
 \label{psconcdyn}
\dot \phi +  \nabla_{i} \left( \phi v_i +  \phi (1-\phi) w_{i}\right) -
   d_{ij} \nabla_i \nabla_{j}(\mu_1 - \bar \mu_2) - \phi(1-\phi)
   d_{ij}^{\,(T)} \nabla_{j}\nabla_i T
 &=& 0 \quad  \\
\label{psconcdyn2}  {\rm or} \quad\quad
\dot \rho_{1} + \bm{v}_{1}\cdot \bm{\nabla} \rho_{1} + \rho_{1} {\rm div}
\bm{v}_{1}
- \rho \, d_{ij} \nabla_i  \nabla_j (\mu_1 - \bar \mu_2) - \frac{\rho_1\rho_2}
{\rho}  d_{ij}^{\,(T)} \nabla_i
 \nabla_j T &=& 0 \quad \\
\label{psconcdyn3}  {\rm or} \quad\quad
\dot \rho_{2} + \bm{v}_{2}\cdot \bm{\nabla} \rho_{2} + \rho_{2} {\rm div}
\bm{v}_{2}
+ \rho \, d_{ij} \nabla_i \nabla_j (\mu_1 - \bar \mu_2) + \frac{\rho_1\rho_2}
{\rho} d_{ij}^{\,(T)}
\nabla_i\nabla_j T &=& 0 \quad
\end{eqnarray}
the entropy dynamics (heat conduction equation)
\begin{equation}
\label{pssigmadyn}
\dot \sigma + v_{i} \nabla_{i} \sigma + \frac{\beta}{\rho} \nabla_i (\rho_1
\rho_2
w_i)
-  \kappa_{ij} \nabla_{i} \nabla_{j} T
-
\frac{\rho_1\rho_2}{\rho} d_{ij}^{\,(T)} \nabla_i \nabla_{j} (\mu_1 - \bar
\mu_2)
= 0
\end{equation}
the elasticity dynamics
\begin{eqnarray}
\label{elast}
\dot U_{ij}\!\! &+&\!\! v_k^{(2)} \nabla_k U_{ij} - \frac{1}{2}(\nabla_j
v_i^{(2)} + \nabla_i
v_j^{(2)}) -\frac{\rho_1}{2}( w_i \nabla_j + w_j \nabla_i) \ln \frac{\rho_2}{\rho}
+ U_{ki} \nabla_j v_k^{(2)} + U_{kj} \nabla_i v_k^{(2)} \nonumber \\&+&\!\!
 \zeta_{l} \delta_{ij}\Phi_{kk} + \zeta_{tr} (\Phi_{ij} - \frac{1}{3} \delta_{ij}
\Phi_{kk}) - \xi_1 \delta_{ij} \Delta \Phi_{kk} - \xi_2 \Delta \Phi_{ij} - \xi_3
(\nabla_i \nabla_j \Phi_{kk} + \delta_{ij} \nabla_k \nabla_l \Phi_{kl})
\nonumber \\ &-&\!\! \xi_4
(\nabla_i \nabla_k   \Phi_{jk} + \nabla_j \nabla_k \Phi_{ik}) =0
\end{eqnarray}
There are nonlinear couplings to the concentration variable (the cubic one has
been
suppressed), which are not
possible in a 1-fluid description.
\\  For
the momentum balance of the two different species we get
\begin{eqnarray}
 \label{psg1dyn}
&&\!\!\!\!\!\!\!\!\!\!\!\!\!\! \rho_{1} \dot v_{i}^{\,(1)} + \rho_{1}
v_{j}^{\,(1)} \nabla_{j} v_{i}^{\,(1)}
+ \frac{\rho_1}{\rho} \nabla_i (p +\frac{1}{2} \rho_{2} (\bm{v}_{1}^{2} -
\bm{v}_{2}^{2})) + \frac{\rho_1 \rho_2}{\rho} \nabla_i
(\mu_1 - \bar \mu_2) +\frac{\rho_1}{\rho} \Phi_{kj} \nabla_i U_{kj}
\nonumber \\ &-&\!\!\!\!  \rho_1
\Phi_{ij} \nabla_j \ln \frac{\rho_2}{\rho} +  \frac{\rho_1 \rho_2}{\rho}
\beta \nabla_i T
 +  \xi_{ij} \rho_1 \rho_2 w_j - \nu_{ijkl}^{\,(1)}
\nabla_j \nabla_l v_k^{\,(1)} - \nu_{ijkl}^{\,(12)}
\nabla_j \nabla_l v_k^{\,(2)} = 0
\\
\label{psg2dyn}
&&\!\!\!\!\!\!\!\!\!\!\!\!\!\! \rho_{2} \dot v_{i}^{\,(2)} + \rho_{2}
v_{j}^{\,(2)} \nabla_{j} v_{i}^{\,(2)}
+ \frac{\rho_2}{\rho} \nabla_i (p - \frac{1}{2} \rho_{1} (\bm{v}_{1}^{2}-
\bm{v}_{2}^{2})) - \frac{\rho_1 \rho_2}{\rho} \nabla_i
(\mu_1 - \bar \mu_2) -  \frac{\rho_1}{\rho} \Phi_{kj}  \nabla_i U_{kj}
\nonumber  \\  &+&\!\!\!\!  \rho_1 \Phi_{ij} \nabla_j \ln \frac{\rho_2}{\rho}
- \frac{\rho_1 \rho_2}{\rho}
\beta \nabla_i T -  \nabla_j \Phi_{ij}
+ \nabla_j ( \Phi_{jk} U_{ik}
+  \Phi_{ik} U_{jk}) -
 \xi_{ij} \rho_1 \rho_2 w_j \nonumber \\ &-&\!\!\!\! \nu_{ijkl}^{\,(2)}
\nabla_j \nabla_l v_k^{\,(2)} - \nu_{ijkl}^{\,(12)}
\nabla_j \nabla_l v_k^{\,(1)} = 0
\end{eqnarray}
Note that although we made the approximation that the linear
elastic
stress does only act on fluid 2, there are inevitably nonlinear contributions
to
fluid 1, too. There is also a (nonlinear) coupling of fluid 1 to the
concentration, if elastic distortions are present.

The different approximations for the viscosities are discussed in
the Appendix.

In order to facilitate actual calculations we also give
eqs.(\ref{psg1dyn},\ref{psg2dyn}) as dynamic equations for the total momentum
and for the relative velocity
\begin{eqnarray}\label{psgdyn}
\rho \dot v_{i}\!\! &+& \!\! \nabla_{i}p + \nabla_{j}\Bigl( \rho v_{i}v_{j} +
\frac{\rho_1 \rho_2}{\rho} w_{i}w_{j} \Bigr) -\nabla_j \Phi_{ij}
\nonumber  \\
&+&\!\! 2\nabla_j (\Phi_{jk} U_{ik})
- \nu_{ijkl} \nabla_{j}\nabla_{l}v_{k} - \frac{\rho_1 \rho_2}{\rho}
\nu^{\,(c)}_{ijkl}\nabla_{j}\nabla_{l}
w_{k}) = 0 \quad \\ \label{pswdyn}
\dot w_{i} \!\!&+&\!\! \Bigl( v_{j} + \frac{\rho_2 - \rho_1}{\rho} w_{j}
\Bigr) \nabla_{j}w_{i} +
\nabla_{i}\Bigl(\mu_{1} - \bar \mu_{2} + \bm{v \cdot w} +
\frac{\rho_2 - \rho_1}{2\rho} \bm{w}^{2} \Bigr) + \rho \xi_{ij} w_{j}
+ \nabla_j \frac{1}{\rho_2} \Phi_{ij}  \nonumber
\\
&+&\!\!\frac{1}{\rho_2} \Phi_{kj} \nabla_i U_{kj}
- \frac{2}{\rho_2} \nabla_j (\Phi_{kj} U_{ik})
- \frac{\rho_1 \rho_2}{\rho} \nu_{ijkl}^{\,(m)}
\nabla_{l}\nabla_{j}w_{k} - \nu^{\,(c)}_{ijkl}\nabla_{j}\nabla_{l} v_{k}
= 0
\end{eqnarray}

In order to
conserve the global incompressibility condition for all times, i.e.
${\rm div} \dot{\bm{v}} =0$, the pressure has to fulfill the relation
\begin{eqnarray}
\label{plaplacep}
\Delta p &=& -\nabla_{i} \nabla_j \,(\rho_{1} v_{i}^{(1)} v_{j}^{(1)} +
\rho_{2} v_{i}^{(2)} v_{j}^{(2)} ) +\nabla_i \nabla_j \Phi_{ij}
- \nabla_i \nabla_j\, (\Phi_{kj} U_{ik} +
\Phi_{ik} U_{jk})\nonumber \\ &&
+  \nu_{ijkl}  \nabla_i \nabla_j \nabla_l  v_k +
\rho_1 \rho_2 \rho^{-1} \nu_{ijkl}^{\,(c)} \nabla_i \nabla_j \nabla_l  w_k
\end{eqnarray}
In contrast to 1-fluid descriptions for simple fluids, where the
incompressibility condition leads to a considerable mathematical
simplification, this is no longer the case for a 2-fluid description due the
complicated form of (\ref{plaplacep}), even if incompressibility is a very good
approximation in physical terms. In particular, $\Delta p$ is not only connected
to compressions ($U_{kk}$), but also to shear deformations, even in linear order.

Of the statics (\ref{psusT}--\ref{Phitrace}) only the following equations remain
\begin{eqnarray}
 \label{pssusT}
\delta T &=& T C_{V}^{{-1}} \, \delta \sigma + \alpha_{\phi}^{-1} \delta \phi
+\alpha_3^{-1} U_{kk}
\\ \label{pssusmu}
\delta (\mu_{1} - \bar \mu_{2}) &=& \rho^{-1} \kappa_{\phi}^{-1} \, \delta
\phi + \rho^{-1} \alpha_{\phi}^{-1} \, \delta \sigma + \kappa_u^{-1} U_{kk}
\\ \label{Phistrace}
\Phi_{kk} &=& c_l U_{kk} + \alpha_3^{-1} \delta \sigma + \rho^{-1} \kappa_{u}^{-1}
 \delta \phi
\end{eqnarray}
with $\delta \phi = \rho^{-1} \delta \rho_{1} = - \rho^{-1} \delta \rho_{2}$,
while eq.(\ref{Phidev}) remains unchanged.
Note that $\delta \mu$ is not needed, but follows from $\delta p$ via eq.
(\ref{pDuhem2}).
\\[0.5cm]

\section{Discussion \label{discussion}}

Within the general framework of hydrodynamics and thermodynamics we have set up a consistent nonlinear 2-fluid description of complex fluids, in particular for lyotropic nematic liquid crystals and polymer solutions or swollen elastomers. Such a general theory determines the frame for any ad-hoc model, which has to be a special case of the general one.  The comparison with the general theory also reveals implicit and explicit assumptions, approximations and possible generalizations of a given model. A simple or "natural" choice in a given model may not be mandatory, but rather imply a presumption. 

Quite generally we find that neither the velocity, with which a certain variable is convected, nor the stress division between the different fluids can be determined by general principles, but is rather system or material dependent. On the other hand, there are certain restrictions and interrelations among the convective velocities and other physical effects that limit the possible choices. For the two densities $\rho_1$, $\rho_2$ e.g.,  the natural choice for the convection velocities seems to be their native velocities $\bm{v}_1$ and $\bm{v}_2$, respectively. This implies that the total density is convected with the mean velocity $\bm{v}$ (as required by mass transport), while the concentration $\phi$ is convected with $(1/\rho)(\rho_{2} \bm{v}_{1}+ \rho_{1} \bm{v}_{2})$. Another obvious choice would be the mean velocity as convection velocity for both, the total density as well as the concentration implying that also $\rho_1$ and $\rho_2$ are convected with $\bm{v}$. However, the actual convection velocity depends on the value of the material dependent (reactive) flow parameters $\gamma_{\perp}$ and $\gamma_{\,\parallel}$, defined in eq. (\ref{phirev}). 

For the nematic degree of freedom the convective velocity again depends on a material parameter ($\beta_1$ defined in eq.(\ref{nrev})) and is not necessarily equal to $\bm{v}_2$ (if fluid 2 is the nematogen). However, the value of $\beta_1$ influences also the flow alignment of the director (and the back flow due to director reorientation), which can  be measured in shear flow experiments. In the case of visco-elastic and elastic media, which are described by a dynamic equation for the (Eulerian) strain tensor $U_{ij}$, there are two velocities involved. One is the usual convection velocity ($v_k \nabla_k U_{ij}$) and the other one occurs in the  "lower convected" part ($U_{kj}\nabla_{i}v_{k} + U_{ki} \nabla_{j}v_{k}$). There is no fundamental reason for the two to be equal and their actual value depends on the (reactive) flow parameters $\beta_6$ and $\beta_7$, respectively, defined in eqs.(\ref{Urev}, \ref{pwrev}).  

For the evolution equations of the momenta special care has to be taken to get a description, which is compatible with general laws (cf. Chapter \ref{stressdiv}). The currents and quasi-currents that enter the description in terms of either the total momentum and the velocity difference or the two individual momenta are not the same as seen in eqs.(\ref{X1allg}, \ref{X2allg}).  In the nematic case the stress division problem depends on the flow alignment parameters as well as on the convection velocity of the director, while in the visco-elastic case the crucial material parameter $\lambda^{(U)}$, eq.(\ref{Urev},\ref{pwrev}), is not related to a convective velocity. The delicate question of viscosities, and approximations related to them, is discussed in detail in the appendix.

A prominent feature of the 2-fluid description is the coupling of the concentration dynamics to the velocity difference. This leads to a frequency dependent effective diffusion and thermo-diffusion, as well as a frequency dependent coupling to the nematic or the visco-elastic degree of freedom. For low frequencies these contributions to the concentration dynamics constitute additional dissipation channels, while for the short-time dynamics (below the relaxation time of the velocity difference) they are reactive.

Recently, 2-fluid descriptions of diffusion in polymeric systems have been given \cite{grmela1, grmela2} based on the GENERIC approach making use of Poisson brackets. A detailed comparison with these formulations is beyond the scope of this manuscript and will be discussed elsewhere. 
\vspace*{0.5cm}

\section*{Acknowledgments}

This research was supported in part by the National Science Foundation under
Grant
No. PHY99-07949.

\vspace*{0.5cm}

\section*{Appendix \label{appendix}}
\setcounter{equation}{0}
\renewcommand{\theequation}{A.\arabic{equation}}

In this Appendix we discuss viscosity and viscosity-like phenomena in the
2-fluid hydrodynamics. We show that in order to get the 1-fluid limit (the
binary liquid) correctly, some care has to be taken when the usual
approximations are made.

If there is only one velocity present, the viscous contribution to the
dissipation function is $
\nu_{ijkl}(\nabla_j v_i)(\nabla_l v_k)$
with $\nu_{ijkl}= \nu_{jikl}=\nu_{ijlk}=\nu_{klij}$, which ensures that only
symmetric velocity gradients contribute to dissipation. Antisymmetric velocity
gradients, $\curl \bm{v}$, describe rotations. A
solid body rotation ($\curl \bm{v}=const.$), however must not increase the
entropy and $(\curl \bm{v})^2$ contributions are not allowed in the dissipation
function. With these symmetries the viscosity tensor has 2 coefficients
for the isotropic 
\begin{equation}\label{nuiso}
\nu_{ijkl}= \nu (\delta_{jl} \delta_{ik} + \delta_{il} \delta_{jk} -
\frac{2}{3}  \delta_{ij} \delta_{kl}) + \zeta  \delta_{ij} \delta_{kl}
\end{equation}
and 5 for the nematic case
\begin{eqnarray} \label{nunema}
\nu_{ijkl}&\!\!=\!\!&  \nu_2 \,(\delta_{jl} \delta_{ik} + \delta_{il} \delta_{jk})
+2(\nu_1 +\nu_2 - 2\nu_3)\,n_i n_j n_k n_l
 + (\nu_5 -\nu_4 + \nu_2)
(\delta_{ij} n_k n_l + \delta_{kl} n_i n_j) \nonumber \\ &&+ (\nu_4 -
\nu_2)\, \delta_{ij}
\delta_{kl}
+ (\nu_3 - \nu_2) (n_j n_l \delta_{ik} + n_j n_k \delta_{il} + n_i n_k
\delta_{jl} + n_i n_l \delta_{jk})
\end{eqnarray}

In a 2-fluid description the same restrictions hold with respect to the mean
velocity $\bm{v}$, since it is the conjugate to the momentum density and
$\curl \bm{v}=const.$ still describes solid body rotations. There are no such
restrictions to the relative velocity $\bm{w}$, and $\curl \bm{m}$ can
contribute to the dissipation. The most general form for viscous dissipation
in a 2-fluid description thus reads
\begin{eqnarray} \label{Rvis}
2R^{(vis)}&=& \nu_{ijkl}(\nabla_j v_i)(\nabla_l v_k) + 2\nu_{ijkl}^{\,(c)}
 (\nabla_j v_i)(\nabla_l m_k) +
\nu_{ijkl}^{\,(w)} (\nabla_j m_i)(\nabla_l m_k) \nonumber \\ &+&
\nu^{\,(r)}_{ij} (\curl\bm{m})_i (\curl \bm{m})_j + 2\nu^{\,(d)}_{ijk}
(\curl \bm{m})_i \nabla_j v_k + 2\nu^{\,(e)}_{ijk}
(\curl \bm{m})_i \nabla_j m_k \quad\quad
\end{eqnarray}
Note that only $\nu_{ijkl}$ has the dimension of a viscosity, while
$\nu_{ijkl}^{\,(c)}$ and $\nu^{(d)}_{ijk}$ are kinematic viscosities, while
$\nu_{ijkl}^{\,(w)}$, $\nu^{(e)}_{ijk}$, and $\nu^{\,(r)}_{ij}$ are
viscosities divided by $\rho^{2}$. The tensors $\nu$ and $\nu^{\,(w)}$ have
the familiar form (\ref{nuiso}) or (\ref{nunema}). For $\nu^{\,(c)}$ there is
no a-priori reason for a $\nu_{ijkl}^{\,(c)}= \nu_{klij}^{\,(c)}$ symmetry,
since $\bm{v}$ and
$\bm{m}$ are not equivalent. However, as will be seen below, a consistent
2-fluid description is only possible, if this symmetry holds and
$\nu_{ijkl}^{\,(c)}$ has the form (\ref{nuiso}) or
(\ref{nunema}).\footnote{Without this symmetry
the form
(\ref{nuiso}) still applies for the isotropic case, while in the nematic case an additional
coefficient is present, i.e. the term $\nu_5 (n_i n_j
\delta_{kl}+n_k n_l \delta_{ij})$ in (\ref{nunema})
splits into two different parts, $\nu_{5a}n_i n_j \delta_{kl} + \nu_{5b}n_k
n_l \delta_{ij}$.} The tensor $\nu^{\,(r)}_{ij}=\nu^{(r)}\delta_{ij}$ or
$\nu^{\,(r)}_{ij}=\nu_{1}^{(r)}\delta_{ij}+ \nu_{2}^{(r)} n_i n_j$ contains 1 or 2
coefficients in the isotropic and nematic case, respectively. The 3rd rank
material tensors, symmetric in the last two indices $\nu^{\,(d,e)}_{ijk}=
\nu^{\,(d,e)}_{ikj}$ are zero in the
isotropic case and both carry
one coefficient in the nematic case $\nu^{\,(d,e)}_{ijk}= \nu^{(d,e)}(
\epsilon_{ikl}n_j n_l + \epsilon_{jkl}n_i n_l)$. For the dissipative currents
this leads to
\begin{eqnarray}\label{gvis}
\sigma_{ij}^{\,(dis)} &=&  -
\nu_{ijkl}\,
\nabla_l v_k - \nu_{ijkl}^{\,(c)} \,\nabla_l m_k
- \nu_{kji}^{(d)} (\curl \bm{m})_{k} \\ \label{wvis}
X_i^{\,(dis)} &=&  \xi_{ij}^\prime \,m_j - \nabla_j
\Bigl( \nu_{ijkl}^{\,(w)}\, \nabla_l m_k + \nu_{klij}^{\,(c)}\, \nabla_l v_k
+ \nu_{kl}^{(r)} \epsilon_{kji} (\curl \bm{m})_{l} + \nu_{lpk}^{(d)}
\epsilon_{lji} \nabla_{p} v_{k} \nonumber \\
&&\,\,\,\,\,\,\,\,\,\,\,\,\,\,\,\,\,\,\,\,\,\,\,\,\,\,\,\,
+\, \nu_{lpk}^{(e)}
\epsilon_{lji} \nabla_{p} m_{k} + \nu_{kji}^{(e)} (\curl \bm{m})_{k}
\Bigr) \quad \quad
\end{eqnarray}

Since there is already friction $\sim m_{i}$, very often the viscosity-like
dissipation $\sim \nabla_{j} m_{i}$ is neglected altogether
($\nu^{(c)}=\nu^{(d)}=\nu^{(e)}= \nu^{(w)}=\nu^{(r)}=0$). Such an
approximation leads to (\ref{ndiss},\ref{gdiss}). In the strong coupling
limit, where $\bm{w}$ (and $\bm{m}$) vanish, this approximation seems to be
appropriate and it correctly gives the 1-fluid limit of binary mixtures. On
the other hand, for two fluids only gently coupled there is no a-priori
reason, why e.g. the tensor $\nu^{(c)}$ (or $\nu^{(w)}$) should be neglected
compared
to $\nu$,
since both terms contain gradients of $\bm{v}_{1}$ as well as of
$\bm{v}_{2}$. Indeed, the dissipation
function in terms of $\bm{v}_{1,2}$ reads
\begin{eqnarray}\label{Rvis2}
2 R^{(vis)} &=&  \nu_{ijkl}^{\,(1)}(\nabla_j v_i^{(1)})(\nabla_l v_k^{(1)})
+ 2 \nu_{ijkl}^{\,(12)}
 (\nabla_j v_i^{(1)})(\nabla_l v_k^{(2)})   +
\nu_{ijkl}^{\,(2)} (\nabla_j v_i^{(2)})(\nabla_l v_k^{(2)})
 \nonumber \\ &+&   \frac{\rho_{1}^{2}\rho_{2}^{2}}{\rho^{2}}
\nu^{\,(r)}_{ij} (\curl[\bm{v}_{1} - \bm{v}_{2}])_{i}
(\curl[\bm{v}_{1} - \bm{v}_{2}])_{j}  \nonumber  \\ &+&
  2 \frac{\rho_{1}\rho_{2}}{\rho} (\curl[\bm{v}_{1} -
\bm{v}_{2}])_{i} ( \nu_{ijk}^{(d1)}\nabla_{j} v_{k}^{(1)}
+ \nu_{ijk}^{(d2)}\nabla_{j} v_{k}^{(2)})
 \end{eqnarray}
where $\curl \bm{v}$ is absent in the dissipation function. Comparing (\ref{Rvis})
and (\ref{Rvis2}) in harmonic approximation, i.e. neglecting cubic
and quartic terms involving e.g. $(\curl \bm{v}_{1})_{i} v_{k}^{(1)} \nabla_{j}
\rho_{1}$,
$(\bm{v}_{1} \times \bm{\nabla} \rho_{1})_{i} v_{k}^{(1)}
\nabla_{j}\rho_{1}$, or
$v_{i}^{(1)}(\nabla_{j}\rho_{1})(\nabla_{l}v_{k}^{(1)})$,
$v_{i}^{(1)}v_{k}^{(1)}(\nabla_{j}\rho_{1})(\nabla_{l}
\rho_{1})$,
we get
\begin{eqnarray}
\label{nud1}
\nu^{(d1)} &=& \rho_{1} \nu^{(d)} + \rho_{1}\rho_{2} \nu^{(e)} \\
\label{nud2}
\nu^{(d2)} &=& \rho_{2} \nu^{(d)} - \rho_{1}\rho_{2} \nu^{(e)} \\
\label{nu1}
\rho^{2}\nu_{ijkl}^{(1)} &=& \rho_{1}^{2} \nu_{ijkl} + 2 \rho_{1}^{2}
  \rho_{2}\nu_{ijkl}^{(c)} + \rho_{1}^{2} \rho_{2}^{2}
\nu_{ijkl}^{(w)}
\\ \label{nu2}
\rho^{2}\nu_{ijkl}^{(2)} &=& \rho_{2}^{2} \nu_{ijkl} - 2 \rho_{2}^{2}
  \rho_{1} \nu_{ijkl}^{(c)} + \rho_{1}^{2} \rho_{2}^{2}
\nu_{ijkl}^{(w)}
\\ \label{nu12}
\rho^{2} \nu_{ijkl}^{(12)} &=&  \rho_{1}\rho_{2} \nu_{ijkl} +
\rho_{1}\rho_{2}(\rho_{2} \nu_{klij}^{(c)} - \rho_{1}
\nu_{ijkl}^{(c)}) -  \rho_{1}^{2} \rho_{2}^{2} \nu_{ijkl}^{(w)}
\end{eqnarray}
Again there is no a-priori reason for $\nu_{ijkl}^{(12)}=
\nu_{klij}^{(12)}$. However, since the tensors $\nu$, $\nu^{(w)}$,
$\nu^{(1)}$, and $\nu^{(2)}$ do have this symmetry, eqs.(\ref{nu1},\ref{nu2})
force $\nu^{(c)}$ to have it, and finally (\ref{nu12}) requires also
$\nu^{(12)}$ to have this symmetry and thus to be of the form (\ref{nuiso},\ref{nunema}).

Neglecting the curl-terms means the same in both descriptions (\ref{Rvis})
and (\ref{Rvis2}), i.e. $\nu^{(r)}= \nu^{(d)}=\nu^{(d1)}=\nu^{(d2)}=0$. For
the symmetric velocity
gradient terms the approximation $\nu^{(c)}= \nu^{(w)}=0$ used in
(\ref{gdiss},\ref{wdiss}) (i.e. no $\nabla_{i}m_{j}$-terms in (\ref{Rvis})) leads to
$\rho_{2}^{2} \nu^{(1)}= \rho_{1}^{2} \nu^{(2)}= \tfrac{1}{2} \rho_{1}\rho_{2}
  \nu^{(12)}$ leaving only
one viscous tensor independent. Neglecting only the cross-viscosity
${\nu^{(c)}}$ in (\ref{Rvis}) does not imply the cross-viscosity in
(\ref{Rvis2}) to vanish, since $\nu^{(c)}=0$ gives $\rho \nu^{(12)}=
\rho_{2} \nu^{(1)} + \rho_{1} \nu^{(2)}$. The opposite case $\nu^{(12)}=0$ used
in (\ref{sg1dyn},\ref{sg2dyn}) leads to a non-zero $\rho_{1}\rho_{2}\nu^{(c)}=
\rho_{2} \nu^{(1)}-
\rho_{1}\nu^{(2)}$ (and $\nu = \nu^{(1)}+
\nu^{(2)}$, $\nu^{(w)}= \rho_{1}^{-2} \nu^{(1)} + \rho_{2}^{-2}
\nu^{(2)}$). Thus, the approximations leading to (\ref{gdiss},\ref{wdiss}) are
not compatible to those used in (\ref{sg1dyn},\ref{sg2dyn}).

In the general case (\ref{Rvis2}) leads to the following viscous contributions
to the left hand sides of (\ref{sg1dyn}) and (\ref{sg2dyn}), respectively
\begin{eqnarray}\label{vis1}
-\nu_{ijkl}^{(1)}\nabla_{j}\nabla_{l}v_{k}^{(1)}
-\nu_{ijkl}^{(12)}\nabla_{j}\nabla_{l}v_{k}^{(2)}
-\frac{\rho_{1}\rho_{2}}{\rho} \nu_{kij}^{(d1)} \nabla_{j}
(\curl[\bm{v}_{1}- \bm{v}_{2}])_{k} \nonumber \\
- \frac{\rho_{1}^{2}\rho_{2}^{2}}{\rho^{2}} \nu^{(r)}_{kl} \epsilon_{kji}\nabla_{j}
(\curl[\bm{v}_{1}-\bm{v}_{2}])_{l}  -
\frac{\rho_{1}\rho_{2}}{\rho} \epsilon_{lji}
\nabla_{j} \bigl(\nu_{lpk}^{(d1)} \nabla_{p}v_{k}^{(1)}+ \nu_{lpk}^{(d2)}
\nabla_{p}v_{k}^{(2)}\bigr)
\end{eqnarray}
and
\begin{eqnarray}\label{vis2}
-\nu_{ijkl}^{(2)}\nabla_{j}\nabla_{l}v_{k}^{(2)}
-\nu_{klij}^{(12)}\nabla_{j}\nabla_{l}v_{k}^{(1)}
-\frac{\rho_{1}\rho_{2}}{\rho} \nu_{kij}^{(d2)} \nabla_{j}
(\curl[\bm{v}_{1}- \bm{v}_{2}])_{k} \nonumber \\
+\frac{\rho_{1}^{2}\rho_{2}^{2}}{\rho^{2}} \nu^{(r)}_{kl} \epsilon_{kji}\nabla_{j}
(\curl[\bm{v}_{1}- \bm{v}_{2}])_{l}  +
\frac{\rho_{1}\rho_{2}}{\rho} \epsilon_{lji}
\nabla_{j} \bigl(\nu_{lpk}^{(d1)} \nabla_{p}v_{k}^{(1)}+ \nu_{lpk}^{(d2)}
\nabla_{p}v_{k}^{(2)}\bigr)
\end{eqnarray}
In the 1-fluid limit $\nu^{(d1,d2)}$ have to vanish and $\rho (\nu^{(1)}+\nu^{(12)})
\to \rho_{1} \nu$ and $\rho(\nu^{(2)}+\nu^{(12)})
\to \rho_{2} \nu$, which is obtained for vanishing $\nu^{(w)}$ and $\nu^{(c)}$.

\vspace*{3cm}

\renewcommand{\baselinestretch}{0.6}
\renewcommand{\refname}{{\bf \Large References\\}}


\begin{thebibliography}{10}


\bibitem{larsonbook}
R.G. Larson, {\it The Structure and Rheology of Complex Fluids},
Oxford University Press (1999).


\bibitem{reichl}
L.E. Reichl, {\it A Modern Course in Statistical Physics},
2nd Ed., Wiley (1997).

\bibitem{drewbook}
D.A. Drew and S.L. Passman, {\it Theory of multi-component fluids},
(Applied Mathematical Sciences, Vol. 135),
Springer Verlag (1998).

\bibitem{martin}
P.C. Hohenberg and P.C. Martin,
{\it Annals Phys.}, {\bf 281} 636 (2000).


\bibitem{park}
L.E. Sugiyama and W. Park,
{\it Phys.Plasmas}, {\bf 7}, 4644 (2000).


\bibitem{lee}
D.K.K. Lee and P.A. Lee,
{\it Physica B}, {\bf 261}, 481 (1999).


\bibitem{hebraud}
P. Hebraud, F. Lequeux, and J.F. Palierne,
{\it Langmuir}, {\bf 16}, 8296 (2000).

\bibitem{kadoma}
I.A. Kadoma and J.W. van Egmond,
{\it Phys. Rev. Lett}, {\bf 80}, 5679 (1998).

\bibitem{lhuillier}
D. Lhuillier,
{\it J. Non-Newtonian Fluid Mech.}, {\bf 96}, 19 (2001).

\bibitem{brochard1}
F. Brochard  and P.G. deGennes,
{\it Macromolecules}, {\bf 10}, 1157 (1977).

\bibitem{harden}
J.L Harden, H. Pleiner, and P.A. Pincus,
{\it J. Chem. Phys.}, {\bf 94}, 5208 (1991).

\bibitem{brochard2}
F. Brochard  and P.G. deGennes,
{\it PhysicoChemical Hydrodynamics}, {\bf 4}, 313 (1983).

\bibitem{doi1}
M. Doi, in {\it Dynamics and Patterns in Complex Fluids.
New Aspects of the Physics-Chemistry Interface},
eds.\ A. Onuki and K. Kawasaki, Springer Proceedings in Physics, Vol. 52,
Springer Verlag, Berlin, (1990).

\bibitem{onuki1}
A. Onuki,
{\it Phys. Rev. Lett}, {\bf 62}, 2472 (1989).

\bibitem{milner1}
S.T. Milner,
{\it Phys. Rev. Lett}, {\bf 66}, 1477 (1991).

\bibitem{milner2}
S.T. Milner
{\it Phys. Rev.E}, {\bf 48}, 3674 (1993).

\bibitem{fredrickson}
E. Helfand and G.H. Fredrickson
{\it Phys. Rev. Lett}, {\bf 62}, 2468 (1989).

\bibitem{helfand}
H. Ji and E. Helfand, {\it Macromolecules}, {\bf 28}, 3869 (1995).

\bibitem{hashimoto}
S. Saito, A. Takenaka, N. Toyoda, and T. Hashimoto,
{\it Macromolecules}, {\bf 34}, 6461 (2001).

\bibitem{doi2}
M. Doi and A. Onuki,
{\it J. Phys. (France)}, {\bf 112}, 1631 (1992).

\bibitem{onuki2}
A. Onuki, R. Yamamoto, and T. Taniguchi,
{\it J. Phys. II (France)}, {\bf 7}, 295 (1997).

\bibitem{okuzono}
T. Okuzono,
{\it Phys. Rev.E}, {\bf 56}, 4416 (1997).

\bibitem{jasnow}
T. Sun, A.C. Balazs, and D. Jasnow,
{\it Phys. Rev.E}, {\bf 59}, 603 (1999).

\bibitem{tanaka}
T. Araki, and H. Tanaka,
{\it Macromolecules}, {\bf 34}, 1953 (2001).

\bibitem{marrucci1}
G. Ianniruberto, and G. Marrucci,
{\it J. Non-Newtonian Fluid Mech.}, {\bf 54}, 231 (1994).

\bibitem{marrucci2}
G. Ianniruberto, F. Greco, and G. Marrucci,
{\it Ind. Eng. Chem. Res.}, {\bf 33}, 2404 (1994).

\bibitem{beris1}
M.V. Apostolakis, V.G. Mavrantzas, and A.N. Beris,
{\it J. Non-Newtonian Fluid Mech.}, {\bf 102}, 409 (2002).

\bibitem{mbuch}
H.R. Brand and H. Pleiner, Hydrodynamics and Electrohydrodynamics of Liquid
Crystals, in {\it Pattern Formation in Liquid Crystals},
eds.\ A. Buka and L. Kramer, Springer New York, p.\ 15-67 (1995)

\bibitem{lubensky}
T.C. Lubensky and P.M. Chaikin,
{\it Principles of Condensed Matter Physics},
Cambridge University Press (1995)

\bibitem{Forster}
D. Forster, {\it Hydrodynamic Fluctuations, Broken Symmetry and Correlation
Functions}, Benjamin, Reading Mass. (1975)

\bibitem{deGennes}
P.G. de Gennes and J. Prost, {\it The Physics of Liquid Crystals},
Oxford University Press (1993)

\bibitem{mpp}
P.C. Martin, O. Parodi, and P.S. Pershan,
{\it Phys. Rev.} {\bf A6}, 2401 (1972).

\bibitem{leslie}
H. Pleiner and H.R. Brand, {\it Continuum Mech. Thermodyn.} {\bf 14}, 297 (2002),

\bibitem{Temmen}
H. Temmen, H. Pleiner, M. Liu, and H.R. Brand, {\it Phys.Rev.Lett.} {\bf 84},
3228 (2000)

\bibitem{pleiner}
H. Pleiner, M. Liu, and H.R. Brand, {\it Rheol. Acta} {\bf 39}, 560 (2000)

\bibitem{grmela1}
A. Elafif, M. Grmela, and G. Lebon, {\it J. Non-Newtonian Fluid Mech.} {\bf 86}, 253 (1999)

\bibitem{grmela2}
A. El Afif and M. Grmela, {\it J. Rheol.} {\bf 46}, 591 (2002)

\end{thebibliography}
\end{document}